\documentclass[12 pt,onecolumn,draftclsnofoot]{IEEEtran}
\usepackage{setspace}
\usepackage[inline]{enumitem}
\usepackage{xcolor}
\usepackage{amsmath,amsfonts}
\usepackage{algorithm}
\usepackage[group-separator={,},group-minimum-digits=4]{siunitx}
\usepackage{threeparttable}
\usepackage{array}
\usepackage[caption=false,font=normalsize,labelfont=sf,textfont=sf]{subfig}
\usepackage{textcomp}
\usepackage{stfloats}
\usepackage{url}
\usepackage{verbatim}
\usepackage{graphicx}
\usepackage{algpseudocode}
\usepackage{cite}
\usepackage{balance}
\usepackage{mathrsfs}

\begin{document}

\title{
{\fontsize{21pt}{\baselineskip}\selectfont Disco Intelligent Reflecting Surfaces: Active Channel Aging for Fully-Passive Jamming Attacks}
}
\author{{\fontsize{11 pt}{\baselineskip}\selectfont
        Huan~Huang,
        Ying~Zhang,~\textit{Student~Member,~IEEE,}
        Hongliang~Zhang,~\textit{Member,~IEEE,}
        Yi~Cai,~\textit{Member,~IEEE,}
        A.~Lee~Swindlehurst,~\textit{Fellow,~IEEE,}
        and~Zhu~Han,~\textit{Fellow,~IEEE}}
\vspace{-0.7cm}
\thanks{
H.~Huang and Y.~Cai are with Jiangsu Engineering Research Center of Novel Optical Fiber Technology and Communication Network, Suzhou Key Laboratory of Advanced Optical Communication Network Technology, Soochow University, Suzhou, Jiangsu 215006, China (e-mail: hhuang1799@gmail.com, yicai@ieee.org).

Y.~Zhang is with the School of Information and Communication Engineering, University of Electronic Science and Technology of China, Chengdu 611731, China (e-mail: yzhang1@std.uestc.edu.cn).

H.~Zhang is with the School of Electronics, Peking University, Beijing 100871, China (email: hongliang.zhang92@gmail.com).

A.~L.~Swindlehurst is with the Center for Pervasive Communications and Computing, University of California, Irvine, CA 92697, USA (e-mail: swindle@uci.edu).

Z. Han is with the Department of Electrical and Computer Engineering at the University of Houston, Houston, TX 77004 USA (email: hanzhu22@gmail.com).} 
}
\maketitle
\vspace{-0.8cm}

\begin{abstract}
\vspace{-0.1cm}
Due to the open communications environment in wireless channels, wireless networks are vulnerable to jamming attacks. However, existing approaches for jamming rely on knowledge of the legitimate users' (LUs') channels, extra jamming power, or both. 
To raise concerns about the potential threats posed by illegitimate intelligent reflecting surfaces (IRSs), we propose an alternative method to launch jamming attacks on LUs without either LU channel state information (CSI) or  jamming power.
The proposed approach employs an adversarial IRS with random phase shifts, referred to as a ``disco" IRS (DIRS), that acts like a ``disco ball" to actively age the LUs' channels.
Such active channel aging (ACA) interference  can be used to launch jamming attacks on multi-user multiple-input single-output (MU-MISO) systems. The proposed DIRS-based fully-passive jammer (FPJ) can jam LUs with no additional jamming power or knowledge of the LU CSI, and it can not be mitigated by classical anti-jamming approaches.
A theoretical analysis of the proposed DIRS-based FPJ that provides an evaluation of the DIRS-based jamming attacks is derived. Based on this detailed theoretical analysis, some unique properties of the proposed DIRS-based FPJ can be obtained.
Furthermore, a design example of the proposed DIRS-based FPJ based on one-bit quantization of the IRS phases is demonstrated to be sufficient for implementing the jamming attack.
In addition, numerical results are provided to show the effectiveness of the derived theoretical analysis and the jamming impact of the proposed DIRS-based FPJ. 
\end{abstract}
\maketitle
\vspace{-0.2cm}
\begin{IEEEkeywords}
\vspace{-0.1cm}
Jamming attacks, intelligent reflecting surface, multi-user MISO (MU-MISO), channel aging, low-power wireless networks.
\end{IEEEkeywords}

\IEEEpeerreviewmaketitle
\newpage
\section{Introduction}
Due to the intrinsically open communications environment in wireless channels, wireless networks are vulnerable to malicious attacks, and it is difficult to protect transmit signals from eavesdroppers~\cite{PLSsur1,DoSsur1,AntiJammingSurv}. To protect the confidentiality of transmit wireless signals, cryptographic techniques are used to prevent eavesdroppers from intercepting transmit information~\cite{crypbook}. Secure communications using cryptographic techniques rely on the computational difficulty of the underlying mathematical process required to unravel the codes, and thus secure communications can be achieved unless the eavesdroppers have extensive computing capabilities.

However, malicious attacks such as jamming (also known as DoS-type attacks in the related literature~\cite{DoSJamm,DoSJamm2}), are more easily implemented than eavesdropping. Generally, jamming attacks can be launched by an active attacker, i.e., an active jammer (AJ), which inflicts intentional interference in order to block the communication between the base station (BS) and the legitimate users (LUs). As discussed below, physical-layer AJs can be divided into the following categories~\cite{DoSsur1}: constant AJs, intermittent AJs, reactive AJs, and adaptive AJs.

\begin{enumerate}
\item Constant AJs: A constant AJ continuously broadcasts jamming signals, such as pseudorandom noise or modulated Gaussian waveforms~\cite{DoSsur1}. However, the energy efficiency of a constant AJ is very low. In practice, energy constraints are an inherent drawback of AJs~\cite{DoSJamm,DoSJamm2}. As a result, the other jamming methods listed below have been investigated~\cite{intermittentAJ,reactiveAJ,adaptiveAJ}.

\item Intermittent AJs: As its name implies, an intermittent AJ only transmits jamming signals from time to time~\cite{intermittentAJ}. The jamming effectiveness of the intermittent AJ is limited since it may or may not be active at the time of communication between the BS and LUs.

\item Reactive AJs: To improve the jamming effectiveness, the reactive AJ approach has been proposed, in which jamming attacks are launched only when communication between the BS and LUs is detected~\cite{reactiveAJ}. As a result, the jamming effectiveness of a reactive AJ is higher than both constant and intermittent AJs.

\item Adaptive AJs: An adaptive AJ achieves the highest jamming effectiveness by adjusting its jamming power according to the time-varying wireless channels between the BS and the LUs~\cite{adaptiveAJ}.
For example, an adaptive AJ achieves better efficiency by reducing jamming power during deep fades or outages in the BS-LU channel, but this requires updated information about the BS-LU channel which is often difficult to obtain.
Typically, an adaptive AJ is used as an idealized approach for benchmarking purposes~\cite{DoSsur1}.
\end{enumerate}

Considering the inherent energy constraint of AJs, \emph{can jamming attacks be launched without jamming power}? Thanks to emerging intelligent reflecting surfaces (IRSs)~\cite{IRSsuradd,IRSsur1,IRSsur11,IRSsur2,IRSsur3,IRSdeployment,HuangRIS}, an adversarial IRS-based passive jammer (PJ) without active power transmission has been proposed for single-user systems~\cite{PassJamSU}, which minimizes the received power at the LU by destructively adding the signal reflected from the IRS. However, the channel state information (CSI) of all channels involved must be known at the illegitimate IRS. CSI for the LU is difficult to obtain in practice, especially for an entirely passive IRS~\cite{DaipartI,HuangCSI,HuangDLRIS}.

To acquire the CSI of IRS-aided channels~\cite{AORIS,AORCG}, the receivers (users) instead of the IRS send pilot signals to the transmitter, and the transmitter then estimates the IRS-aided channels using methods such as the least squares (LS) algorithm~\cite{DaipartI}.
In other words, if the illegitimate IRS wants to obtain LU CSI, it needs to perform channel estimation jointly with the legitimate BS and LUs. Although an adversarial IRS can ideally impose a harmful impact on wireless networks~\cite{IIRSSur}, the assumption that the IRS knows the CSI of all channels involved is unrealistic for practical wireless networks.

Consequently, we ask the following research question: \emph{Can jamming attacks be launched without either jamming power or LU CSI}? The authors of~\cite{FPJDiscoIRS} have investigated the downlink of a multi-user MISO (MU-MISO) system jammed by an IRS-based fully-passive jammer (FPJ). The illegitimate IRS-based FPJ can jam LUs without jammer power and CSI by aging all involved channels during both the \emph{pilot transmission (PT)} phase and the \emph{data transmission (DT)} phase.

To raise concerns about the significant potential threats posed by illegitimate IRSs, we propose a disco-IRS-based (DIRS-based) FPJ that can launch jamming attacks on the LUs without relying on either jamming power or LU CSI.
The main contributions$\footnote{A portion of this work was published in~\cite{FPJDiscoIRS}, where we have illustrated the impact of the FPJ on the downlink of an MU-MISO system using only simulations and not a theoretical analysis.}$ are summarized as follows:  
\begin{itemize}
\item We investigate the uplink of an MU-MISO system jammed by the proposed DIRS-based FPJ, which is the first jamming attack proposed that can be launched without jamming power or LU CSI. Before the~\emph{DT} phase, channel estimation is performed by the legitimate system during the~\emph{PT} phase to provide the CSI for designing the decoder, during which time the illegitimate DIRS remains silent$\footnote{The term ``silent" means that the wireless signals are perfectly absorbed by the illegitimate IRS~\cite{RISSilent}.}$. 
    The DRIS is then activated during the \emph{DT} phase, and the DIRS phase shifts are randomly generated. The DIRS with random phase shifts acts like a ``disco ball'' distributing the BS energy in random directions. Consequently, the BS-LU channels change rapidly, and serious interference, referred to as active channel aging (ACA) interference, is introduced. Since random IRS reflection coefficients are employed, there is no need for the DIRS to know the LU CSI.
\item We perform a theoretical analysis of our proposed DIRS-based FPJ for cases where the BS uses the zero-forcing (ZF) or maximum-ratio combining (MRC)  detector. More specifically, lower bounds on the ergodic achievable uplink rates achieved in the presence of the proposed DIRS-based FPJ are determined, which provides an evaluation of the jamming effectiveness of the proposed approach. The simulation results show that the derived lower bounds are close to the obtained ergodic achievable uplink rates.
\item Based on the detailed theoretical analysis, we present some unique properties of the proposed DIRS-based FPJ as follows: 
    1) In contrast to AJs, the jamming impact of the proposed DIRS-based FPJ cannot be mitigated by increasing the transmit power; 2) The jamming impact of the proposed DIRS-based FPJ is not dependent on the quantization nor the distribution of the discrete random DIRS phase shifts.
\item We show that the proposed DIRS-based FPJ can be implemented with reflecting elements whose individual phase shifts are determined by a single bit.
    Since the jamming impact is based on ACA interference introduced by the proposed DIRS, the characteristics of ACA interference, for instance, the carrier frequency and the bandwidth, are the same as the LUs' transmit signals. As a result, classic anti-jamming technologies such as frequency hopping are not valid for the proposed DIRS-based FPJ.
\end{itemize}

The rest of this paper is organized as follows. In Section~\ref{Princ}, the uplink of an MU-MISO system jammed by the proposed DIRS-based FPJ is modeled, and the performance metric used to quantify the jamming impact is given. Moreover, some useful results on random variables are presented.
In Section~\ref{LowerBound}, the theoretical analysis of the proposed DIRS-based FPJ is performed. Then, some properties of the proposed DIRS-based FPJ are obtained based on the derived theoretical analysis.
Simulation results are provided in Section~\ref{ResDis} to show the effectiveness of the derived theoretical analysis and the performance of the proposed DIRS-based FPJ. Finally, the main conclusions are given in Section~\ref{Conclu}.

\emph{Notation:} In this work, we employ bold capital letters for a matrix, e.g., $\bf{W}$, lowercase bold letters for a vector, e.g., $\boldsymbol{w}_k$, and italic letters for a scalar, e.g., $N_t$. The superscripts $(\cdot)^{H}$ and $(\cdot)^{T}$ denote the Hermitian transpose and the transpose, respectively. Moreover, the symbols $|\cdot|$ and $\|\cdot\|$ denote the absolute value and the Frobenius norm, respectively. 
\section{System Model and Preliminaries}\label{Princ}
In Section~\ref{DIRSFPJ}, we illustrate the uplink of an MU-MISO system jammed by the proposed DIRS-based FPJ and give the general model of the DIRS-based jamming attacks. In Section~\ref{IIB}, we state the performance metric used to quantify the jamming impact of the proposed DIRS-based FPJ. The channel model is presented in Section~\ref{ChannModel}. In Section~\ref{RandomVar}, some important results on random variables are presented, which are used for the theoretical analysis in Section~\ref{LowerBound}.

\subsection{Disco-IRS-Based Fully-Passive Jammer}\label{DIRSFPJ}
\begin{figure}[!t]
\centering
\includegraphics[scale=0.82]{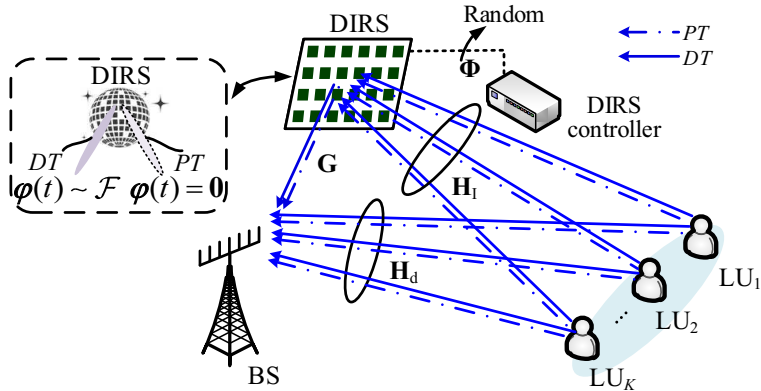}
\caption{The uplink of an MU-MISO system jammed by the disco-IRS-based fully-passive jammer (DIRS-based FPJ). During the \emph{pilot transmission (PT)} phase, the DIRS is silent; During the \emph{data transmission (DT)} phase, the phase shifts of the DIRS reflecting elements are randomly generated by the independent DIRS controller.}
\label{fig1}
\end{figure}

Fig.~\ref{fig1} schematically illustrates the uplink of an MU-MISO system jammed by the proposed DIRS-based FPJ, where the DIRS-based jamming attacks on the LUs are launched without relying on either jamming power or LU CSI. The BS is equipped with an ${N_t}$-element uniform linear array (ULA) and communicates with $K$ legitimate single-antenna users denoted by ${\rm {LU}}_{1}, {\rm {LU}}_{2}, \cdots, {\rm {LU}}_{K}$. A DIRS comprised of $N_{\rm D}$ reflecting elements is deployed near the BS$\footnote{Many existing performance-enhancing IRS-aided systems assume that legitimate IRSs are deployed close to users in order to maximize system performance~\cite{IRSdeployment}. However, in the jamming scenario presented here, we make the more robust assumption that the independent DIRS controller does not have any information about the LUs, such as the LUs' locations and CSI. The location of the BS is fixed, and thus we assume that the DIRS is deployed near the BS. Our deployment strategy is informed by the conclusion given in~\cite{IRSdeployment}, which makes the impact of the DIRS as large as possible. In addition, the distance from the DIRS to the legitimate BS should be greater than the minimum channel correlation distance to ensure that the DIRS-based channel and the direct channel of each LU are independent.}$ to launch jamming attacks on the LUs. 

\underline{\textit{Pilot Transmission:}} The LUs' CSI is obtained in the \emph{PT} phase~\cite{PT,CSIRIS3} in order for the BS to design a decoder used during the \emph{DT} phase, such as the ZF linear detector~\cite{LDetector1}. In particular, during the \textit{PT} phase, the independent DIRS controller acts to make the DIRS absorb the wireless signals it receives. 
The pilot signal used by ${\rm{LU}}_k$ is denoted by ${ {s}_{{\rm{p}},k}}$. 
Consequently, the received pilot vector ${\boldsymbol{y}_{{\rm{p}},k}} \in \mathbb{C}^{{N_t}\times 1}$ at the BS is given by
\begin{equation}
{\boldsymbol{y}_{{\rm{p}},k}} = \sqrt {{p_{{\rm{p}},k}}}{\boldsymbol{h}_{{\rm{d}},k}}{s_{{\rm{p}},k}} + {\boldsymbol{n}_{{\rm{p}}}},
\label{PTeq}
\end{equation}
where ${p_{{\rm{p}},k}}$ is the power of the pilot signal sent by ${\rm{LU}}_k$, and ${\boldsymbol{h}_{{\rm{d}},k}}\in \mathbb{C}^{{N_t}\times 1}$ represents the direct channel between ${\rm{LU}}_k$ and the BS during the \textit{PT} phase. In addition, ${\boldsymbol{n}_{{\rm{p}}}} = {\left[ {{n_{{\rm{p}},1}},{n_{{\rm{p}},2}}, \cdots, {n_{{\rm{p}},{N_t}}}} \right]^T}$ denotes the receiver noise vector at the BS which is assumed to be composed of independent and identically distributed (i.i.d.) elements with zero mean and variance $\sigma_{\rm p}^2$, i.e., ${{n}_{{\rm{p}},i}} \sim \mathcal{CN}\left(0,\sigma_{\rm p}^2\right), i = 1,2, \cdots ,{N_{t}}$.

Based on the received pilot signal ${\boldsymbol{y}_{{\rm{p}},k}}$, the direct channel ${\boldsymbol{h}_{{\rm{d}},k}}$ can be estimated by the BS.
Similarly, the multi-user direct channel ${{\bf{H}}_{\rm d}}$ between BS and all LUs can be obtained, where ${{\bf{H}}_{\rm d}} = \left[ {{\boldsymbol{h}_{{\rm{d}},1}},{\boldsymbol{h}_{{\rm{d}},2}}, \cdots ,{\boldsymbol{h}_{{\rm{d}},K}}} \right] \in \mathbb{C}^{{N_t}\times {K}}$.
We assume that perfect CSI is obtained by the BS during the \textit{PT} phase~\cite{CSIRIS3}, as imperfect CSI is not a core concern in the jamming attack scenario, and the impact of imperfect CSI has also been well studied~\cite{LDetector1,ImprefectCSI}.

\underline{\textit{Linear Detector Design:}}
Two common approaches for decoding at the BS are the MRC and ZF detectors.
The conventional MRC detector is given by
\begin{equation}
{{\bf{W}}^H} = {\bf{H}}_{\rm{d}}^H = {\left[ {{\boldsymbol{w}_1},{\boldsymbol{w}_2}, \cdots ,{\boldsymbol{w}_K}} \right]^H}.
\label{MRCeq}
\end{equation}
On the other hand, the conventional ZF detector is 
\begin{equation}
{{\bf{W}}^H} = {\left( {{\bf{H}}_{\rm{d}}^H{{\bf{H}}_{\rm{d}}}} \right)^{ - 1}}{\bf{H}}_{\rm{d}}^H = {\left[ {{\boldsymbol{w}_1},{\boldsymbol{w}_2}, \cdots ,{\boldsymbol{w}_K}} \right]^H}.
\label{ZFeq}
\end{equation}

\underline{\textit{Data Transmission:}} During the \textit{DT} phase$\footnote{Similar to the assumption of reactive AJs in~\cite{PTDTSyn}, a minimum reaction period is required to perform channel sensing and jamming initialization. Therefore, during the PT phase, the pilot signal could be transmitted without being jammed.}$, the LUs transmit their data using the same time-frequency resource as the pilot data. Meanwhile, the DIRS controller tunes the illegitimate IRS to randomly generate phase shifts. Namely, the reflecting vector $\boldsymbol{\varphi}(t) \!= \!\left[ {{e^{j{\varphi _1}(t)}},{e^{j{\varphi _2}(t)}}, \cdots ,{e^{j{\varphi _{{N_{\rm{D}}}}(t)}}}} \right]$ has random phase shifts, i.e., $\varphi_r(t) \sim {\cal F}\left([0,2\pi]\right), r = 1,2,\cdots, N_{\rm D}$.
Since $\left[ {\varphi _1}, {\varphi _2}, \cdots ,{\varphi _{{N_{\rm{D}}}}} \right]$ are randomly generated, the DIRS controller does not use CSI to optimize $\boldsymbol{\varphi}(t)$.

In the proposed DIRS-based FPJ, the DIRS reflecting vector $\boldsymbol{\varphi}(t)$ is time-varying for the \emph{PT} phase and the \emph{DT} phase. Therefore, the illegitimate IRS with random phase shifts acts like a ``disco ball", as shown in Fig.~\ref{fig1}. 
Consequently, the multi-user DIRS-jammed channel between the BS and the LUs is expressed as
\begin{equation}
{{\bf{H}}_{\rm{D}}} = {{\bf{G}}^H}{\rm{diag}}\left( \boldsymbol{\varphi}(t)  \right){{\bf{H}}_{\rm{I}}} = \left[ {{\boldsymbol{h}_{{\rm{D}},1}},{\boldsymbol{h}_{{\rm{D}},2}}, \cdots ,{\boldsymbol{h}_{{\rm{D}},K}}} \right],
\label{BSIRSchanneq}
\end{equation}
where ${\bf{G}}$ represents the channel between the BS and the DIRS, ${{\bf{H}}_{\rm{I}}} = \left[ {{\boldsymbol{h}_{{\rm{I}},1}},{\boldsymbol{h}_{{\rm{I}},2}}, \cdots ,{\boldsymbol{h}_{{\rm{I}},K}}} \right]$ represents the multi-user channel between the DIRS and the LUs, and ${\boldsymbol{h}_{{\rm{D}},k}}$ represents the DIRS-jammed channel between the BS and ${\rm {LU}}_k$ and can be written as ${\boldsymbol{h}_{{\rm{D}},k}}= {{\bf{G}}^H}{\rm{diag}}\left( \boldsymbol{\varphi}(t)\right){{\boldsymbol{h}}_{{\rm{I}},k}}$.

The vector ${\boldsymbol{y}_{\rm{d}}}$ received at the BS is then expressed by
\begin{equation}
{\boldsymbol{y}_{\rm{d}}} = {\sqrt {{p_{\rm{d}}}}} {{\bf{W}}^H}\left( {{{\bf{H}}_{\rm{D}}} + {{\bf{H}}_{\rm{d}}}} \right){{\boldsymbol{s}}_{\rm d}} + {{\bf{W}}^H}{\boldsymbol{n}_{\rm{d}}},
\label{receivedsigeq}
\end{equation}
where ${{p_{\rm{d}}}}$ is the average transmit power of each LU during the \textit{DT} phase, and the vector of transmit symbols ${{\boldsymbol{s}}_{\rm d}} $ during the \textit{DT} phase is given by ${{\boldsymbol{s}}_{\rm d}}= {\left[ {{s_{{\rm{d}},1}},{s_{{\rm{d}},2}}, \cdots, {s_{{\rm{d}},{K}}}} \right]^T}$ where ${s_{{\rm{d}},{k}}}$ denotes the symbol transmitted by ${\rm{LU}}_k$. The receiver noise vector ${\boldsymbol{n}_{{\rm{d}}}}$ during the \textit{DT} phase has zero mean independent elements with variance $\sigma_{\rm d}^2$, and is written as ${\boldsymbol{n}_{{\rm{d}}}}= {\left[ {{n_{{\rm{d}},1}},{n_{{\rm{d}},2}}, \cdots, {n_{{\rm{d}},{N_t}}}} \right]^T}$, i.e., ${{n}_{{\rm{d}},n}} \sim \mathcal{CN}\left(0,\sigma_{\rm d}^2\right), n = 1,2, \cdots ,{N_{t}}$.
\subsection{Ergodic Achievable Uplink Rate}\label{IIB}
The achievable uplink rate of the symbol transmitted by ${\rm {LU}}_k$ is written as ${R_{{\rm d},k}} = {{{\log }_2}\left( {1 + {\gamma _k}} \right)}$, where $\gamma_k$ represents the received  signal-to-interference-plus-noise ratio (SINR) of the $k$-th transmit symbol $s_{{\rm d},k}$. Based on \eqref{receivedsigeq}, the $k$-th element of the received vector ${\boldsymbol{y}_{\rm{d}}}$ is expressed by
\begin{equation}
{y_{{\rm{d}},k}} = \underbrace {\sqrt {{p_{\rm{d}}}} {\boldsymbol{w}_k^H}{\boldsymbol{h}_{{\rm{d}},k}}{s_{{\rm{d}},k}}}_{{\rm{signal}}} + \underbrace {\sqrt {{p_{\rm{d}}}} \sum\limits_{i \ne k,i = 1}^K {{\boldsymbol{w}_k^H}{\boldsymbol{h}_{{\rm{d}},i}}{s_{{\rm{d}},i}}} }_{{\mathop{{\rm{inter-user}}\;{\rm{interference}}}}} + \underbrace {\sqrt {{p_{\rm{d}}}} \sum\limits_{i = 1}^K {{\boldsymbol{w}_k^H}{\boldsymbol{h}_{{\rm{D}},i}}{s_{{\rm{d}},i}}} }_{{\rm{ACA}}\;{\rm interference}} + \underbrace {{{\boldsymbol{w}_k^H}}{\boldsymbol{n}_{\rm{d}}}}_{{\rm{noise}}},
\label{kthsymeq}
\end{equation}
where the ACA interference is caused by the DIRS. Note that even if the DIRS-jammed channel ${\bf{H}}_{\rm D}$ is fixed, it can not be useful because the legitimate BS only knows the aged ${\bf{H}}_{\rm D}$.

When the MRC detector is used, the ergodic rate ${\left. {{{\overline R }_{{\rm{d}},k}}} \right|_{{\rm{MRC}}}}$ is expressed as
\begin{alignat}{1}
\nonumber
{\left. {{{\overline R }_{{\rm{d}},k}}} \right|_{{\rm{MRC}}}}  &= \mathbb{E}\left[ {{{\log }_2}\left( {1  +  {\left. {{\gamma _k}} \right|_{{\rm{MRC}}}}} \right)} \right]\\
&= \mathbb{E}\left[ {{{\log }_2}\left( {1 + \frac{{{p_{\rm{d}}}{{\left\| {{\boldsymbol{h}_{{\rm{d}},k}}} \right\|}^4}}}{{{p_{\rm{d}}}\sum\limits_{i \ne k,i = 1}^K {{{\left| {\boldsymbol{h}_{{\rm{d}},k}^H{\boldsymbol{h}_{{\rm{d}},i}}} \right|}^2}}  + {p_{\rm{d}}}\sum\limits_{i = 1}^K {{{\left| {\boldsymbol{h}_{{\rm{d}},k}^H{\boldsymbol{h}_{{\rm{D}},i}}} \right|}^2}} + \sigma _{\rm{d}}^2{{\left\| {\boldsymbol{h}_{{\rm{d}},k}^H} \right\|}^2}}}} \right)} \right].
\label{GrgGamakMRCeq}
\end{alignat}
For the ZF detector, the ergodic rate ${\left. {{{\overline R }_{{\rm{d}},k}}} \right|_{{\rm{ZF}}}}$ reduces to
\begin{alignat}{1}
\nonumber
{\left. {{{\overline R }_{{\rm{d}},k}}} \right|_{{\rm{ZF}}}} &= \mathbb{E}\left[ {{{\log }_2}\left( {1  +  {\left. {{\gamma _k}} \right|_{{\rm{ZF}}}}} \right)} \right]\\
&= \mathbb{E}\left[ {{{\log }_2}\left( {1 + \frac{{{p_{\rm{d}}}}}{{{p_{\rm{d}}} \sum\limits_{i = 1}^K {{\left| {{\boldsymbol{w}_k^H}{\boldsymbol{h}_{{\rm{D}},i}}} \right|}^2} } + {{\sigma_{\rm d}^2}{\left\| {{\boldsymbol{w}_k}} \right\|^2}}}} \right)} \right].
\label{GrgGamakZFeq}
\end{alignat}

The introduced ACA interference is somewhat similar to the channel aging (CA) interference caused by imperfect CSI~\cite{ChanAge}. However, they are essentially different. For example, we can prove that ${\overline R _{{\rm{d}},k}}$ under the proposed DIRS-based jamming attack tends to zero as the number of DIRS reflecting elements increases.
In other words, as the DIRS grows in size, the FPJ is able to ultimately prevent the BS from receiving any information transmitted by the LUs even though no jamming power nor LU CSI is exploited.
For more unique properties of the proposed DIRS-based FPJ, please see Section~\ref{LowerBound}.
\subsection{Channel Model}\label{ChannModel}
The DIRS is deployed close to the BS, and therefore we assume the BS-DIRS channel ${\bf{G}}$ follows Rician fading~\cite{IRSsur1,Ricianfading}. On the other hand, we assume both the multi-user direct channel ${\bf{H}}_{\rm{d}}$ and the multi-user DIRS-LU channel ${{\bf{H}}_{\rm{I}}}$ follow Rayleigh fading.
Specifically, the Rician fading channel ${\bf{G}}$ is modeled as~\cite{Ricianfading}

\begin{equation}
{{\bf{G}}} = \left[ { \boldsymbol{g}_1, \boldsymbol{g}_2, \cdots ,\boldsymbol{g}_{{N_t}} } \right]= {\sqrt{{\mathscr{L}}_{\rm G}}}  \left( { {{\bf{G}}^{{\rm{LOS}}}}{\sqrt {{{{\boldsymbol{\mathscr{E}}}}}{\left({{{\boldsymbol{\mathscr{E}}}}}+{\bf{I}}_{N_t}\right)^{-1}} }} + {{\bf{G}}^{{\rm{NLOS}}}}{\sqrt {{\left({{{\boldsymbol{\mathscr{E}}}}}+{\bf{I}}_{N_t}\right)^{-1}}}} } \right),
\label{Ricianchan}
\end{equation}
where ${\mathscr{L}}_{\rm G}$ denotes the geometric attenuation and log-normal shadow fading (the large-scale channel fading) between the BS and DIRS.
Moreover, the $N_{t} \times N_{t}$ diagonal matrix ${\boldsymbol{\mathscr{E}}}={\rm{diag}}\left({\varepsilon _1},{\varepsilon _2},\cdots,{\varepsilon _{N_t}}\right)$ is comprised of the Rician factors, and each Rician factor represents the ratio of signal power in the line-of-sight (LOS) component to the scattered power in the non-line-of-sight (NLOS) component. 

In~\eqref{Ricianchan}, \,${\bf{G}}^{{\rm{LOS}}} \,=\, \left[\, {\boldsymbol{g}_1^{{\rm{LOS}}},\, \boldsymbol{g}_2^{{\rm{LOS}}},\, \cdots ,\,\boldsymbol{g}_{{N_t}}^{{\rm{LOS}}}}\, \right]$\, represents\, the\, LOS\, component\,, and ${\bf{G}}^{{\rm{NLOS}}}\,=\,$ \\$\left[ {\boldsymbol{g}_1^{{\rm{NLOS}}}, \boldsymbol{g}_2^{{\rm{NLOS}}}, \cdots ,\boldsymbol{g}_{{N_t}}^{{\rm{NLOS}}}} \right]$ denotes the NLOS component of ${{\bf{G}}}$. The NLOS component ${\bf{G}}^{{\rm{NLOS}}}$ follows Rayleigh fading, while element $\left[{\bf{G}}^{{\rm{LOS}}}\right]_{rn}$ in the LOS component ${\bf{G}}^{{\rm{LOS}}}$ is modeled as~\cite{AORIS,Ricianfading,IRSsur1},
\begin{equation}
{\left[ {{{\bf{G}}^{{\rm{LOS}}}}} \right]_{rn}} = {e^{j\frac{{2\pi }}{\lambda }d\left( {n - 1} \right)\sin {\theta _r}}}, r=1,2,\cdots,N_{\rm D}, n = 1,2,\cdots, N_t,
\label{RicianchanLOS}
\end{equation}
where $\theta _r \in \left[ { - {\theta _{\rm{A}}},{\theta _{\rm{A}}}} \right]$ ($0 < {\theta _{\rm{A}}} \le \pi $) is the angle of arrival (AoA) from the $r$-th reflecting element, $\lambda$ is the wavelength of the transmit symbols, and $d$ represents the antenna-spacing of the ULA at the BS. Meanwhile, the NLOS component ${\bf{G}}^{{\rm{NLOS}}}$ has i.i.d. elements given by $\left[{\bf{G}}^{{\rm{NLOS}}}\right]_{rn} \sim \mathcal{CN}\left(0,1\right)$.
The multi-user DIRS-LU channel ${{\bf{H}}_{\rm{I}}}$ and the multi-user direct channel ${\bf{H}}_{\rm{d}}$ are represented as
\begin{alignat}{1}
&{{\bf{H}}_{\rm{I}}} = {\widehat {\bf{H}}_{\rm{I}}}{{\bf{D}}_{\rm I}^{{1 \mathord{\left/
 {\vphantom {1 2}} \right.
 \kern-\nulldelimiterspace} 2}}} = \left[ {{\sqrt{{{\mathscr{L}}_{{\rm I},1}}}}{{\widehat {\boldsymbol{h}}}_{{\rm I},1}},{\sqrt{{{\mathscr{L}}_{{\rm I},2}}}}{{\widehat {\boldsymbol{h}}}_{{\rm I},2}}, \cdots ,{\sqrt{{{\mathscr{L}}_{{\rm I},K}}}}{{\widehat {\boldsymbol{h}}}_{{\rm I},K}}} \right], \label{HIkeq}\\
&{{\bf{H}}_{\rm{d}}} = {\widehat {\bf{H}}_{\rm{d}}}{{\bf{D}}_{\rm d}^{{1 \mathord{\left/
 {\vphantom {1 2}} \right.
 \kern-\nulldelimiterspace} 2}}} = \left[ {{\sqrt{{{\mathscr{L}}_{{\rm d},1}}}}{{\widehat {\boldsymbol{h}}}_{{\rm d},1}},{\sqrt{{{\mathscr{L}}_{{\rm d},2}}}}{{\widehat {\boldsymbol{h}}}_{{\rm d},2}}, \cdots ,{\sqrt{{{\mathscr{L}}_{{\rm d},K}}}}{{\widehat {\boldsymbol{h}}}_{{\rm d},K}}} \right],
\label{Hdkeq}
\end{alignat}
where elements of the $K\times K$ diagonal matrices ${\bf{D}}_{\rm I} = {\rm{diag}}\left({{\mathscr{L}}_{{\rm I},1}},{{\mathscr{L}}_{{\rm I},2}},\cdots,{{\mathscr{L}}_{{\rm I},K}}\right)$ and ${\bf{D}}_{\rm d} = {\rm{diag}}\left({{\mathscr{L}}_{{\rm d},1}},{{\mathscr{L}}_{{\rm d},2}},\cdots,{{\mathscr{L}}_{{\rm d},K}}\right)$ model the geometric attenuation and log-normal shadow fading, which are assumed to be independent over $n$~\cite{LDetector1}. The elements of ${\widehat {\bf{H}}_{\rm{I}}}$ and ${\widehat {\bf{H}}_{\rm{d}}}$ are i.i.d. Gaussian variables defined as $\left[{\widehat {\bf{H}}_{\rm{I}}}\right]_{rk},\left[{\widehat {\bf{H}}_{\rm{d}}}\right]_{nk} \sim \mathcal{CN}\left(0,1\right), r=1,2,\cdots, N_{\rm D},\; n=1,2,\cdots, N_t, \;k = 1,2,\cdots,K$.
\subsection{Review of Some Results on Random Variables}\label{RandomVar}
\subsubsection{Jensen's Inequality}
Consider a convex function $f:\;{\cal I} \to \mathbb{R}$, where ${\cal I}$ is an interval in $\mathbb{R}$. If a random variable $x \in {\cal I}$, and $f\left({\mathbb E}\left[x\right]\right)$ and ${\mathbb E}\left[f\left(x\right)\right]$ are finite, then
\begin{equation}
{\mathbb E}\left[f\left(x\right)\right] \ge f\left({\mathbb E}\left[x\right]\right).
\label{JensenIneq}
\end{equation}
\subsubsection{Weak Law of Large Numbers}
Consider the following random vector of i.i.d. Lebesgue integrable random variables: ${\boldsymbol{x}} \buildrel \Delta \over = {\left[ {{x_1},{x_2}, \cdots ,{x_n}} \right]^T}$, where ${\mathbb E}\left[x_1\right]= {\mathbb E}\left[x_2\right] = \cdots = {\mathbb E}\left[x_n\right] = \mu$. The weak law of large numbers states that
\begin{equation}
\overline X = \frac{{\sum\limits_{i = 1}^n {{x_i}} }}{n}\mathop  \to \limits^{\rm{p}} \mu ,\;{\rm{as}}\;n \to \infty.
\label{LLNeq}
\end{equation}
In other words, the sample average $\overline X$ converges in probability towards the expected value $\mu$ as $n \to \infty$.
\subsubsection{Lindeberg-L$\acute{e}$vy Central Limit Theorem}
Suppose the random vector ${\boldsymbol{x}} \buildrel \Delta \over = {\left[ {{x_1},{x_2}, \cdots ,{x_n}} \right]^T}$ is a vector of i.i.d. random variables with mean ${\mathbb E}\left[x_1\right]= {\mathbb E}\left[x_2\right] = \cdots = {\mathbb E}\left[x_n\right] = \mu < \infty$ and variance ${\rm {Var}}\left[x_1\right]= {\rm {Var}}\left[x_2\right] = \cdots = {\rm {Var}}\left[x_n\right] = {\sigma}^2 < \infty$. Then, the Lindeberg-L$\acute{e}$vy central limit theorem states that the random variable $\sqrt n \left( {\overline X  - \mu } \right)$ converges in distribution to $\mathcal{CN}\left( {0,{\sigma ^2}} \right)$ as $n \to \infty$, i.e.,
\begin{equation}
\sqrt n \left( {\overline X  - \mu } \right) = \frac{{\sum\limits_{i = 1}^n {{x_i}} }}{{\sqrt n }} - \sqrt n \mu \mathop  \to \limits^{\rm{d}} \mathcal{CN}\left( {0,{\sigma ^2}} \right),\;{\rm{as}}\;n \to \infty.
\label{CLTeq}
\end{equation}
\section{Ergodic Achievable Uplink Rate Under DIRS-Based Jamming Attacks}\label{LowerBound}
In Section~\ref{LBdrived}, we determine lower bounds on the ergodic rates jammed by the proposed DIRS-based FPJ for the cases where the BS uses MRC and ZF detectors, respectively. Then, in Section~\ref{ProperFPJ}, we present various properties of the proposed DIRS-based FPJ: 1) We compare the proposed DIRS-based FPJ to an IRS-based PJ to show that our approach is able to launch jamming attacks without any jamming power and any LU CSI; 2) In contrast to the AJs, the jamming impact of the proposed DIRS-based FPJ cannot be mitigated by increasing the LU transmit power; 3) 
The jamming impact of the proposed DIRS-based FPJ does not depend on the quantization nor the distribution of the random DIRS.
In Section~\ref{FPJex}, based on these properties, we describe a simple implementation of the proposed DIRS-based FPJ using a one-bit IRS with phase shifts following a uniform distribution.
\subsection{Lower Bound of Ergodic Achievable Uplink Rate}\label{LBdrived}
A lower bound is derived in order to approximate the ergodic rate the BS can achieve under the proposed DIRS-based jamming attacks. The following lower bound can be obtained by using Jensen's inequality in a straightforward way:
\begin{alignat}{1}
\nonumber
{\overline R _{{\rm{d}},k}} &= {\mathbb{E}\left[{R_{{\rm d},k}}\right]} \\
\nonumber
&= {\mathbb{E}\left[ {{{\log }_2}\left( {1 + \frac{1}{{\gamma _k^{ - 1}}}} \right)} \right]} \\
& \ge {{{{\log }_2}\left( {1 +  {\frac{1}{ \mathbb{E}\left[ {\gamma _k^{ - 1}} \right] }} } \right)} }, k = 1,2,\cdots, K.
\label{LBJensen}
\end{alignat}
Unfortunately, the lower bound in~\eqref{LBJensen} is implicit. To this end, we present more-explicit lower bounds on the ergodic rate ${\overline R _{{\rm{d}},k}}$ for the cases where the BS uses the MRC and ZF detectors, given below respectively as Proposition~\ref{Proposition2} and Proposition~\ref{Proposition3}.

When the MRC detector is used at the BS to receive the transmit symbol vector ${\boldsymbol{s}}_{\rm d}$, we can derive the following explicit lower bound on the ergodic rate ${\left. {{{\overline R }_{{\rm{d}},k}}} \right|_{{\rm{MRC}}}}$ ($k = 1,2,\cdots, K$) under the jamming launched by the proposed DIRS-based FPJ.
\newtheorem{theorem}{Proposition}
\begin{theorem}
\label{Proposition2}
The lower bound on the ergodic rate ${\left. {{{\overline R }_{{\rm{d}},k}}} \right|_{{\rm{MRC}}}}$ converges in probability towards a fixed value 
as $N_{\rm D}, N_t \to \infty$, i.e.,
\begin{alignat}{1}
\nonumber
{\left. {{{\overline R }_{{\rm{d}},k}}} \right|_{{\rm{MRC}}}} &\ge {{{{\log }_2} \left( {1 + {\frac{1}{ \mathbb{E}\left[ {{{\left. {\gamma _k^{ - 1}} \right|}_{{\rm{MRC}}}}} \right] }} } \right)} } \\
&\mathop  \to \limits^{\rm{p}} {\log _2} \! \left( \!{1 + \frac{{{p_{\rm{d}}}\left( {{N_t} - 1} \right){\mathscr{L}_{{\rm{d}},k}}}}{{{p_{\rm{d}}}\sum\limits_{i = 1,i \ne k}^K {{\mathscr{L}_{{\rm{d}},i}}}  + {p_{\rm{d}}}{N_{\rm{D}}}\sum\limits_{i = 1}^K {{\mathscr{L}_{\rm{G}}}{\mathscr{L}_{{\rm{I}},i}}}  + \sigma _{\rm{d}}^2}}} \right),\;{\rm{as}}\; N_{\rm D}, N_t  \to \infty.
\label{MRCLBeq}
\end{alignat}
\end{theorem}

\begin{IEEEproof}
See Appendix~\ref{AppendixA}.
\end{IEEEproof}

When the BS uses the ZF detector to receive ${\boldsymbol{s}}_{\rm d}$, the following explicit lower bound on the ergodic rate ${\left. {{{\overline R }_{{\rm{d}},k}}} \right|_{{\rm{ZF}}}}$ ($k = 1,2,\cdots, K$) is presented in Proposition~\ref{Proposition3}.
\begin{theorem}
\label{Proposition3}
The lower bound of the ergodic rate ${\left. {{{\overline R }_{{\rm{d}},k}}} \right|_{{\rm{ZF}}}}$ converges in probability towards a fixed value 
as $N_{\rm D}, N_t \to \infty$, i.e.,
\begin{alignat}{1}
\nonumber
{\left. {{{\overline R }_{{\rm{d}},k}}} \right|_{{\rm{ZF}}}} &\ge {{{{\log }_2} \left( {1 + {\frac{1}{ \mathbb{E}\left[ {{{\left. {\gamma _k^{-1} } \right|}_{{\rm{ZF}}}}} \right] }} } \right)} } \\
&\mathop  \to \limits^{\rm{p}} {\log _2} \! \left( \!{1 + \frac{{{p_{\rm{d}}}\left( {{N_t} - K} \right){\mathscr{L}_{{\rm{d}},k}}}}{{ {p_{\rm{d}}}{N_{\rm{D}}}\sum\limits_{i = 1}^K {{\mathscr{L}_{\rm{G}}}{\mathscr{L}_{{\rm{I}},i}}}  + \sigma _{\rm{d}}^2}}} \right),\;{\rm{as}}\; N_{\rm D}, N_t  \to \infty.
\label{ZFLBeq}
\end{alignat}
\end{theorem}

\begin{IEEEproof}
See Appendix~\ref{AppendixB}.
\end{IEEEproof}

The lower bounds in~\eqref{MRCLBeq} and~\eqref{ZFLBeq} provide accurate estimates of ergodic rates. In Section~\ref{ResDis}, the simulation results show that the derived lower bounds are close to the actual ergodic rates. 
\subsection{Properties of Disco-IRS-Based Fully-Passive Jammer}\label{ProperFPJ}
In this subsection, we illustrate some unique properties of the proposed DIRS-based FPJ to show the difference between it and existing jammers.
\subsubsection{Jamming Users Without Jamming Power and LU CSI}
In Section~\ref{IIB}, we have illustrated that the proposed DIRS-based FPJ jams LUs via the DIRS-based ACA. To the best of our knowledge, this is the first instance of an illegitimate jammer that is able to launch jamming attacks without either jamming power or LU CSI.

Although the IRS-based PJ approach proposed in~\cite{PassJamSU} can launch jamming attacks without jamming power, a very demanding requirement must be met to implement this PJ: the CSI of all channels involved, such as the direct channel and the IRS-aided channels, must be known. The IRS-based PJ proposed in~\cite{PassJamSU} can be extended to MU-MISO systems by implementing the following optimization:
\begin{alignat}{1}
&\mathop {\min }\limits_{\boldsymbol{\varphi}} \sum\limits_{k = 1}^K{R_{{\rm{d}},k}} = \mathop {\min }\limits_{\boldsymbol{\varphi}}  \sum\limits_{k = 1}^K {\log _2}\left( {1 + \frac{{{p_{\rm{d}}}{{\left| {\boldsymbol{w}_k^H\left( {{\boldsymbol{h}_{{\rm{d}},k}} + {\boldsymbol{h}_{{\rm{D}},k}}} \right)} \right|}^2}}}{{p_{\rm{d}}}{\sum\limits_{i = 1,i \ne k}^K {{{\left| {\boldsymbol{w}_k^H\left( {{\boldsymbol{h}_{{\rm{d}},i}} + {\boldsymbol{h}_{{\rm{D}},i}}} \right)} \right|}^2}}  + \sigma _{\rm{d}}^2{{\left\| {{\boldsymbol{w}_k}} \right\|}^2}}}} \right) \label{addeq81}\\
&{\rm{s}}.{\rm{t}}.\;\; \boldsymbol{\varphi}  = \left[ {{e^{j{\varphi _1}}},{e^{j{\varphi_2}}}, \cdots ,{e^{j{\varphi _{{N_{\rm{D}}}}}}}} \right], \label{addeq82}\\
&\;\;\;\; \;\; \;  {\varphi _r}  \in \left[ {0,2\pi } \right], r= 1,2,\cdots, N_{\rm D} \label{addeq83}.
\end{alignat}
The objective function in~\eqref{addeq81} is a continuous and differentiable function of ${{\boldsymbol{\varphi}}}$, and the constraints in~\eqref{addeq82} and~\eqref{addeq83} create a complex circle manifold. Therefore, the above optimization problem can be computed by the Riemannian conjugate gradient (RCG) algorithm~\cite{AORCG,RCGAlg} as follows: generate the Riemannian gradient; determine the search direction; and retract the tangent vector.

(1) Riemannian Gradient:
For ease of presentation, we denote the objective function in~\eqref{addeq81} as
\begin{equation}
{\mathcal{G}\!\left( {\boldsymbol{\varphi}}\right)} = \sum\limits_{k = 1}^K {\log _2}\left( {1 + \frac{{{p_{\rm{d}}}{{\left| {\boldsymbol{w}_k^H\left( {{\boldsymbol{h}_{{\rm{d}},k}} + {\boldsymbol{h}_{{\rm{D}},k}}} \right)} \right|}^2}}}{{p_{\rm{d}}}{\sum\limits_{i = 1,i \ne k}^K {{{\left| {\boldsymbol{w}_k^H\left( {{\boldsymbol{h}_{{\rm{d}},i}} + {\boldsymbol{h}_{{\rm{D}},i}}} \right)} \right|}^2}}  + \sigma _{\rm{d}}^2{{\left\| {{\boldsymbol{w}_k}} \right\|}^2}}}} \right).
\label{RCGreq}
\end{equation}
Consequently, the Riemannian gradient at $ {\boldsymbol{\varphi}}$ is a tangent vector that denotes the greatest decreasing direction of $\mathcal{G}( {\boldsymbol{\varphi}})$, which is given by
\begin{equation}
{\rm{grad}}{\mathcal{G}\!\left({\boldsymbol{\varphi}}\right)} = \nabla\!\mathcal{G}\!\left({\boldsymbol{\varphi}} \right) - {\mathop{\rm Re}\nolimits} \left\{ {\nabla \!\mathcal{G}\!\left({\boldsymbol{\varphi}} \right)\odot{{\boldsymbol{\varphi}} ^H}} \right\}\odot {\boldsymbol{\varphi}},
\label{RCG1}
\end{equation}
where $\nabla\! \mathcal{G}\!\left( {\boldsymbol{\varphi}} \right)$ represents the Euclidean gradient.

(2) Search Direction:
The tangent vector conjugate to ${\rm{grad}}{\mathcal{G}\!\left( {\boldsymbol{\varphi}} \right)}$ can be used as the search direction $\mathcal{D}$, which is expressed as
\begin{equation}
\mathcal{D}= -{\rm{grad}}{\mathcal{G}\!\left({\boldsymbol{\varphi}} \right)} +{\rho_1 (\tilde{\mathcal{D}}-{\mathop{\rm Re}\nolimits} \{ {\tilde{\mathcal{D}}\odot{{\boldsymbol{\varphi}} ^H}} \}\odot {\boldsymbol{\varphi}} ) } ,
\label{RCG2}
\end{equation}
where $\rho_1$ and $\tilde{\mathcal{D}}$ denote the conjugate gradient update parameter and the previous search direction, respectively.

(3) Retraction:  Retract the tangent vector back to the complex circle manifold described by~\eqref{addeq82}. Mathematically,
\begin{equation}
\frac{{{{\left( {{\boldsymbol{\varphi}} + {\rho _2}\mathcal{D}} \right)}_n}}}{{\left| {{{\left( {{\boldsymbol{\varphi}}  + {\rho _2}\mathcal{D}} \right)}_n}} \right|}} \mapsto {{\boldsymbol{\varphi}}_n},
\label{RCG3}
\end{equation}
where $\rho_2$ represents the Armijo step size.

To solve the optimization problem via the RCG algorithm, the IRS controller requires significant computing power.
Moreover, the requirement that the controller has access to the CSI of ${\bf{H}}_{\rm d}$, $\bf{G}$, and ${\bf{H}}_{\rm I}$ is difficult to realize due to the passive nature of the IRS~\cite{DaipartI}. In particular, to acquire the CSI of the IRS-aided channels $\bf{G}$ and ${\bf{H}}_{\rm I}$, the LUs instead of the IRS send pilot signals to the BS, and the BS then estimates the IRS-aided channels by traditional solutions, for instance, the least squares (LS) algorithm~\cite{DaipartI}. When a legitimate IRS is used to enhance system performance, the CSI of the IRS-aided channels can be jointly estimated by the BS, the LUs, and the IRS. However, acquiring the IRS-aided channels at the illegitimate IRS is not realistic in the jamming attack scenario.

The DIRS-based ACA interference is introduced by randomly generating phase shifts for the DIRS. Compared to the IRS-aided PJ, the proposed DIRS-based FPJ does not need significant computing power. On the other hand, the DIRS approach launches jamming attacks without requiring the LU CSI.

\subsubsection{Larger Transmit Power Increases Jamming}
Taking~\eqref{ZFLBeq} as an example, if the LUs increase the average transmit power ${p_{\rm{d}}}$, the term ${p_{\rm{d}}}{N_{\rm{D}}}\sum\nolimits_{i = 1}^K {{\mathscr{L}_{\rm{G}}}{\mathscr{L}_{{\rm{I}},i}}}$ in the denominator also increases. In other words, the jamming attacks launched by the proposed DIRS-based FPJ cannot be mitigated by increasing the average transmit power, since increasing the power leads to even more serious jamming.


Note that an AJ is also able to jam the vector ${\boldsymbol{y}}_{\rm d}$ received by the BS without LU CSI. More specifically, consider a single-antenna AJ that broadcasts the jamming signal symbol ${s}_{\rm J}$ with power $p_{\rm J}$. The $k$-th element of the received vector ${\boldsymbol{y}_{\rm{d}}}$ under active jamming is given by
\begin{equation}
{y_{{\rm{d}},k}} =  \underbrace {\sqrt {{p_{\rm{d}}}} {\boldsymbol{w}_k^H}{\boldsymbol{h}_{{\rm{d}},k}}{s_{{\rm{d}},k}}}_{{\rm{signal}}} + \underbrace {\sqrt {{p_{\rm{d}}}} \sum\nolimits_{i \ne k,i = 1}^K {{\boldsymbol{w}_k^H}{\boldsymbol{h}_{{\rm{d}},i}}{s_{{\rm{d}},i}}} }_{{\mathop{{\rm{inter-user}}\;{\rm{interference}}}}} + \underbrace {\sqrt {{p_{\rm{J}}}} {{\boldsymbol{w}_k^H}{\boldsymbol{h}_{{\rm{J}}}}{s_{{\rm{J}}}}} }_{{\rm{AJ}}\;{\rm interference}} + \underbrace {{{\boldsymbol{w}_k^H}}{\boldsymbol{n}_{\rm{d}}}}_{{\rm{noise}}},
\label{kthsymAJeq}
\end{equation}
where ${\boldsymbol{h}_{{\rm{J}}}}$ denotes the channel between the BS and the AJ whose elements have zero mean and variance ${\mathscr{L}_{\rm{J}}}$, which represents the geometric attenuation and log-normal shadow fading between the BS and the AJ. Specifically, ${\boldsymbol{h}_{\rm{J}}} = \sqrt{\mathscr{L}_{\rm{J}}}{\boldsymbol{\widehat h}_{\rm{J}}}$ and $\left[{\boldsymbol{\widehat h}_{\rm{J}}}\right]_n \sim \mathcal{CN}\left( {0,1} \right), n =1,2,\cdots,N_t$.
Consequently, AJ interference proportional to the jamming power $P_{\rm J}$ is introduced into the achievable rate ${R_{{\rm d},k}}$. Mathematically, the achievable rate $R_{{\rm{d}},k}^{{\rm{AJ}}}$ under active jamming is formulated as
\begin{equation}
R_{{\rm{d}},k}^{{\rm{AJ}}} = {\log _2}\!\left(\! {1 + \frac{{{p_{\rm{d}}}{{\left| {\boldsymbol{w}_k^H{\boldsymbol{h}_{{\rm{d}},k}}} \right|}^2}}}{{{p_{\rm{d}}}\sum\limits_{i = 1,i \ne k}^K {{{\left| {\boldsymbol{w}_k^H{\boldsymbol{h}_{{\rm{d}},i}}} \right|}^2}}  + {p_{\rm{J}}}{{\left| {\boldsymbol{w}_k^H{\boldsymbol{h}_{\rm{J}}}} \right|}^2} + \sigma _{\rm{d}}^2{{\left\| {{\boldsymbol{w}_k}} \right\|}^2}}}} \right).
\label{kthrateAJeq}
\end{equation}
If we take the case where the BS adopts the ZF detector~\eqref{ZFeq} to receive the symbols sent by the LUs, the achievable rate ${\left. {R_{{\rm{d}},k}^{{\rm{AJ}}}} \right|_{{\rm{ZF}}}}$ expressed in~\eqref{kthrateAJeq} reduces to
\begin{equation}
{\left. {R_{{\rm{d}},k}^{{\rm{AJ}}}} \right|_{{\rm{ZF}}}} = {\log _2}\!\left(\! {1 + \frac{{{p_{\rm{d}}} }}{{  {p_{\rm{J}}}{{\left| {\boldsymbol{w}_k^H{\boldsymbol{h}_{\rm{J}}}} \right|}^2} + \sigma _{\rm{d}}^2{{\left\| {{\boldsymbol{w}_k}} \right\|}^2}}}} \right).
\label{kthrateZFAJeq}
\end{equation}

Although the AJ can jam the LUs without their CSI, the jamming attacks launched by the AJ can be mitigated by increasing the average transmit power ${p_{\rm{d}}}$. In order to achieve the desired jamming impact, the AJ has to increase its jamming power ${p_{\rm{J}}}$. However, high-power jamming signals are more easily detected by the legitimate BS, which increases the risk of AJ exposure. 

\subsubsection{Ergodic Achievable Uplink Rate Independent of Precision and Stochastic Distribution of Reflecting Phase Shifts}
Based on~\eqref{MRCLBeq} and~\eqref{ZFLBeq}, an interesting property can be observed: the jamming impact of the proposed DIRS-based FPJ does not depend on the quantization of the IRS phase shifts. In other words, jamming launched by a one-bit DIRS-based FPJ is equivalent to that launched by the proposed FPJ using an infinite-precision DIRS.

In addition, the jamming impact of the proposed DIRS-based FPJ does not depend on the distribution of the random phase shifts. 
As long as the number of DIRS reflecting elements is large, the ergodic rate will tend to zero even if the proposed DIRS-based FPJ is implemented using a one-bit IRS with uniformly distributed (i.e., equally likely) one-bit phase shifts. In practice, it is easy to implement such a simple IRS with a large number of reflecting elements~\cite{RIS256ele,RISCB}. 

An IRS is an ultra-thin surface inlaid with massive sub-wavelength reflecting elements whose electromagnetic responses (such as phase shifts) can be controlled, for example, by simple programmable PIN diodes~\cite{RISCB}. Based on the ON/OFF behavior of the PIN diodes, however, only a  limited set of discrete phase shifts can be achieved by an IRS. Some existing work has investigated the trade-off between performance and the number of bits used to determine the phase shifts~\cite{IRSelement,IRSUPA,DiscreteIRS}. Empirically, the higher the resolution of the discrete IRS phase shifts, the higher the cost but the better the performance.

Taking the IRS-based PJ in~\eqref{addeq81} as an example, the optimization of the phase shifts via the RCG algorithm implicitly assumes continuously tunable phase shifts. Assuming that the illegal IRS has $b$-bit quantized phase shifts, the discrete reflecting phase shifts $ {\widetilde{\boldsymbol{\varphi}}} = \left[{\widetilde\varphi_1}, {\widetilde\varphi_2},\cdots, {\widetilde\varphi_{N_{\rm D}}}  \right]$ must be calculated by finding the quantized values closest to the result of the optimization:
\begin{alignat}{1}
&\mathop {\min }\limits_{ {\widetilde{\boldsymbol{\varphi}}} }  {\left\| {{{\boldsymbol{\varphi}}} - {\widetilde{\boldsymbol{\varphi}}} } \right\|^2} \label{addeq101}\\
&{\rm{s}}.{\rm{t}}.\;\; {\widetilde \varphi_r} \in \left\{ {0,\frac{{2\pi }}{{{2^b}}}, \cdots ,\frac{{2\left( {{2^b} - 1} \right)\pi }}{{{2^b}}}} \right\}, r=1,2,\cdots, N_{\rm D}. \label{discretephs}
\end{alignat}
Obviously, in this approach based on the LU CSI, the higher the quantization resolution, the more serious the jamming impact of the IRS-based PJ. Such a high resolution discrete phase-shift design is unnecessary in our proposed approach.

\subsection{One-Bit DIRS-Based FPJ Nullifying Anti-Jamming Technologies}\label{FPJex}
\begin{figure}[!t]
\centering
\includegraphics[scale = 0.92]{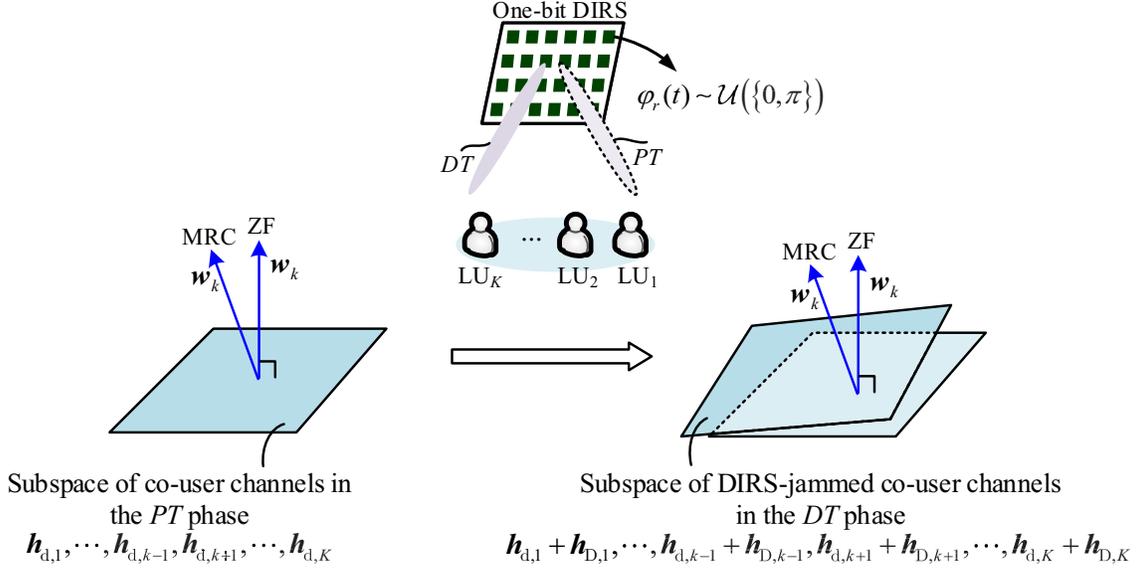}
\caption{One-bit disco-IRS-based fully-passive jammer (DIRS-based FPJ) by actively aging channels to launch jamming attacks on legitimate users (LUs), where the reflecting phase shifts follow a simple one-bit discrete distribution.}
\label{fig2}
\end{figure}

Based on the properties stated in Section~\ref{ProperFPJ}, the proposed DIRS-based FPJ can be implemented by using a one-bit IRS, where the reflecting phase shifts follow a simple discrete uniform distribution, i.e., ${\varphi _r} \sim {\cal U}\left( {\left\{ {0,\pi } \right\}} \right)$. Fig.~\ref{fig2} illustrates the proposed DIRS-based FPJ. By actively aging the wireless channels between the BS and the LUs, serious ACA interference is introduced.

The detector used at the BS is designed based only on the multi-user direct channel ${{\bf{H}}_{\rm{d}}}$. For example, the ZF decoder ${{\boldsymbol{w}}_k}$ is only orthogonal to the subspace of the direct co-user channels $\boldsymbol{h}_{{\rm{d}},1},\cdots,\boldsymbol{h}_{{\rm{d}},k-1},\boldsymbol{h}_{{\rm{d}},k+1},\cdots,\boldsymbol{h}_{{\rm{d}},K}$, as depicted in Fig.~\ref{fig2}, which is different from the DIRS-jammed co-user channel.

From~\eqref{kthsymeq}, the jamming attacks launched by the proposed DIRS-based FPJ are implemented by introducing ACA interference. Since the characteristics of ACA interference, such as the carrier frequency, are the same as the LUs' transmit signals, classic anti-jamming technologies such as frequency hopping~\cite{Frequencyhopping,Frequencyhopping1} are not helpful for the proposed DIRS-based FPJ. 

\section{Simulation Results and Discussion}\label{ResDis}
In this section, we provide numerical results to evaluate the effectiveness of the proposed DIRS-based FPJ in Section~\ref{Princ} and that of the derived theoretical analysis in Section~\ref{LowerBound}. Consider an MU-MISO system with eight single-antenna LUs jammed by the proposed DIRS-based FPJ in Section~\ref{Princ}.
More specifically, the BS is located at (0m, 0m, 2m) and the eight LUs are randomly distributed in the circular region $S$ centered at (0m, 160m, 0m) with a radius of 10m. Furthermore, the DIRS with $N_{\rm D}$ reflecting elements is deployed at ($-d_{\rm{BD}}$m, 0m, 2m) to launch the proposed DIRS-based fully-passive jamming on the LUs. The distance between the BS and the DIRS, the number of DIRS reflecting elements, and the number of antennas at the BS are $d_{\rm{BD}} = 2$, $N_{\rm D} = 4096$, and $N_{t} = 256$.
If not otherwise specified, the numbers of DIRS reflecting elements, antennas, and LUs, as well as the BS-DIRS distance default to these values. The influence of the numbers of reflecting elements, antennas, and LUs as well as that of the BS-DIRS distance are discussed next.
The propagation parameters of wireless channels ${\bf{H}_{\rm D}}$, ${\bf{H}_{\rm I}}$, and ${{\bf{G}}}$ described in Section~\ref{ChannModel} are defined in Table~\ref{tab1} based on 3GPP propagation models~\cite{3GPP}, and the variance of the noise is $\sigma^2_{\rm d}\!=\!-170\!+\!10\log _{10}\left(BW\right)$ dBm.
\begin{figure}[!t]
\centering
\includegraphics[scale=0.40]{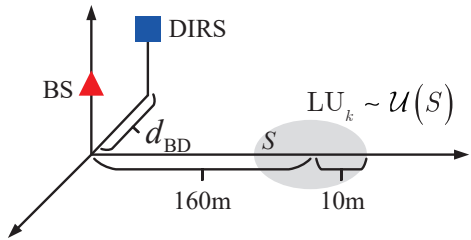}
\caption{A visualization of an MU-MISO system jammed by the proposed DIRS-based FPJ, where the BS is located at (0m, 0m, 2m), the DIRS with $N_{\rm D}$ reflecting elements is deployed at the location ($-d_{\rm{BD}}$m, 0m, 2m), and eight LUs are randomly distributed in the circular region $S$ centered at (0m, 160m, 0m) with a 10m radius.}
\label{fig3}
\end{figure}

In this section, we illustrate the performance of the following benchmarks: the ergodic rates resulting from an MU-MISO system without jamming attacks~\cite{LDetector1}, where the BS adopts the ZF detector (ZF w/o Jamming) or the MRC detector (MRC w/o Jamming); the ergodic rates described in~\eqref{kthrateZFAJeq} resulting from an MU-MISO system jammed by an AJ with -4dBm jamming power (AJ w/ -4dBm) and  4dBm jamming power (AJ w/ 4dBm), where the AJ is deployed at (20m,160m,0m); and the ergodic rates given in~\eqref{GrgGamakMRCeq} and~\eqref{GrgGamakZFeq} resulting from an MU-MISO system jammed by the proposed DIRS-based FPJ, where the BS also adopts the ZF detector (DIRS-FPJ \& ZF) or the MRC detector (DIRS-FPJ \& MRC).

\begin{table}
\footnotesize
\centering
\caption{Wireless Channel Simulation Parameters}
\label{tab1}
\begin{threeparttable}
\begin{tabular}{ c|c|c }
\hline
Parameter        &Notation     &Value\\
\hline
Large-scale LOS channel fading     &${ {\mathscr{L}_{\rm{G}}} }$   & $ 35.6 + 22{\log _{10}}({d}) $ (dB) \\
\hline
Large-scale NLOS channel fading    &${ {\mathscr{L}_{{\rm{d}},k}},{\mathscr{L}_{{\rm{I}},k}}}$    &$32.6+36.7{\log _{10}}({d})$ \\
\hline
Transmission bandwidth                       &$BW$                                                  &180 kHz\\
\hline
Rician factors                       &${\boldsymbol{\mathscr{E}}}$                                 &$10{\bf{I}}_{N_{t}}$ \\
\hline
\end{tabular}
\end{threeparttable}
\end{table}

\subsubsection{Ergodic Achievable Uplink Rate Versus Average Transmit Power}
\begin{figure}[!t]
\centering
\includegraphics[scale=0.75]{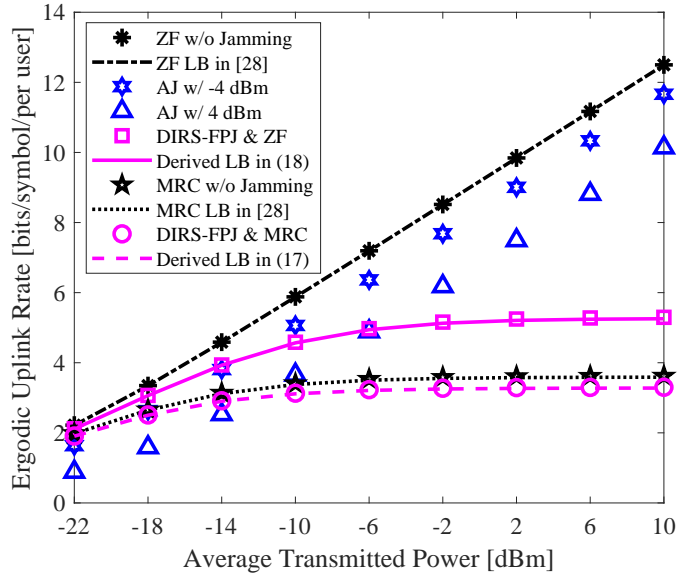}
\caption{Ergodic achievable uplink rate vs average transmit power of each LU for different benchmarks.}
\label{fig4}
\end{figure}

Fig.~\ref{fig4} illustrates the relationship between the ergodic rates and the average transmit power of each LU. The ergodic rates of ZF w/o Jamming,  AJ w/ -4dBm, AJ w/ 4dBm, DIRS-FPJ \& ZF, MRC w/o Jamming, as well as DIRS-FPJ \& MRC benchmarks are depicted, respectively. Meanwhile, the lower bounds of ZF w/o Jamming and MRC w/o Jamming given in~\cite{LDetector1} are shown in Fig.~\ref{fig4}. To show the effectiveness of the theoretical analysis derived in Section~\ref{LowerBound}, the lower bounds of the proposed DIRS-based FPJ are also included in Fig.~\ref{fig4}.

From Fig.~\ref{fig4}, one can see that the proposed DIRS-based FPJ can effectively jam the LUs without either jamming power or LU CSI.
The AJ approach requires a significant amount of extra jamming energy, but increased transmit power at the LUs not only fails to mitigate the jamming impact of the proposed DIRS-based FPJ but even aggravates it. As shown in Fig.~\ref{fig4}, as the average transmit power of each LU increases, the jamming impact of the proposed DIRS-based FPJ gradually becomes stronger and eventually exceeds that of the AJ. Although an MU-MISO system using low-order modulations, such as quadrature phase shift keying (QPSK), can work in the low power domain to reduce DIRS-based ACA interference, we will see below that increasing the number of DIRS reflecting elements can ensure reasonable jamming attacks. Furthermore, if the transmit signal is in the low power domain, a relatively small amount of jamming can provide an effective jamming impact.

Compared to the case where the BS uses the MRC detector, the jamming impact of the proposed DIRS-based FPJ for ZF decoding is stronger, as can be seen from Fig.~\ref{fig4}. In Section~\ref{FPJex}, we have illustrated that the proposed DIRS-based FPJ jams the LUs by introducing ACA interference. However, serious inter-user interference (IUI), which is a type of multi-user interference (MUI), has been introduced when the BS uses the MRC detector to receive the signals transmitted by the LUs. As a result, the jamming impact caused by the ACA interference is suppressed. However, the IUI is well suppressed when the BS employs the ZF detector, and thus the jamming impact of the proposed DIRS-based FPJ is more evident.
\subsubsection{Ergodic Achievable Uplink Rate Versus Number of DIRS Reflecting Elements}
\begin{figure*}[htbp]
\centering
 \begin{minipage}{0.65\linewidth}
     \centering
     \includegraphics[width=0.88\linewidth]{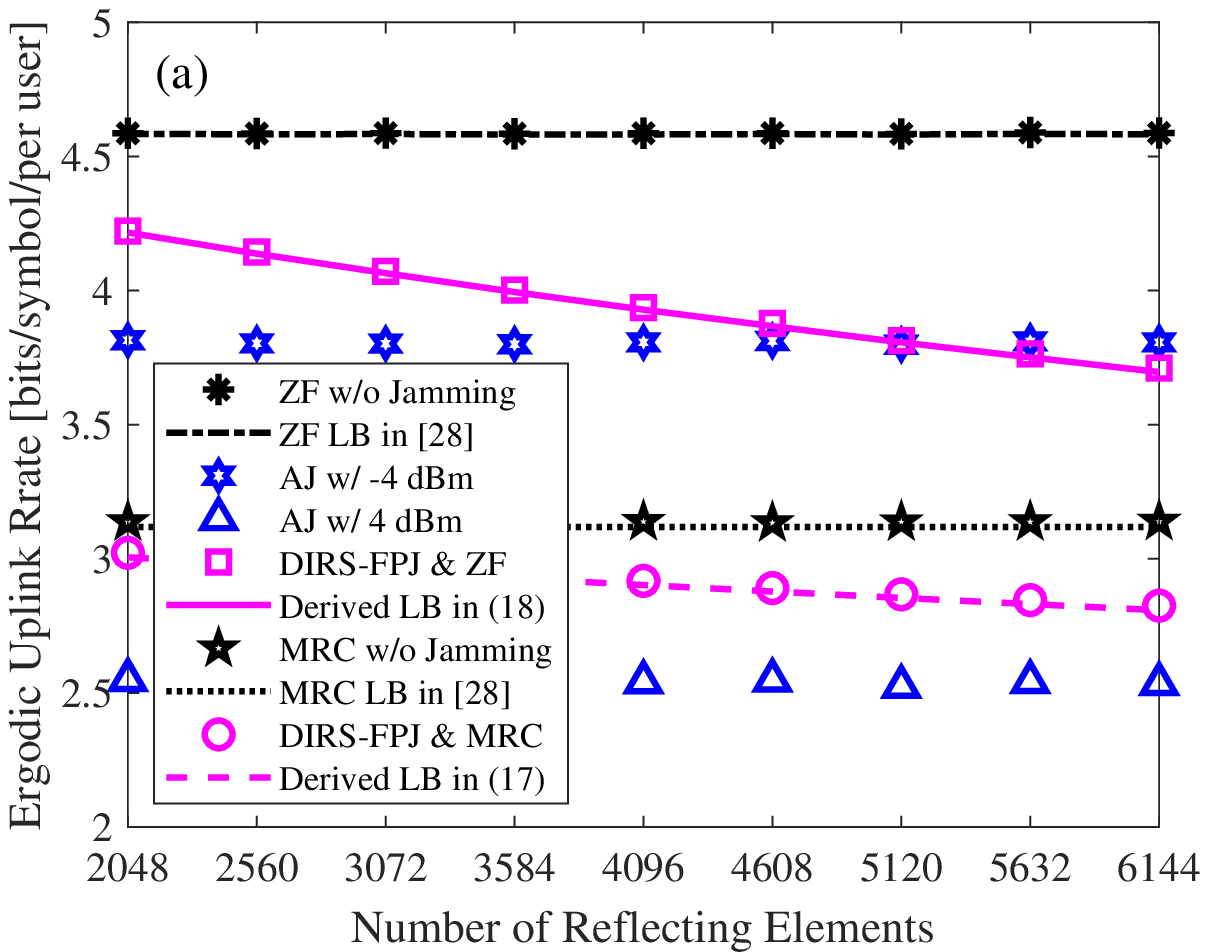}
     \label{fig5a}
 \end{minipage}

    \begin{minipage}{0.65\linewidth}
     \centering
     \includegraphics[width=0.88\linewidth]{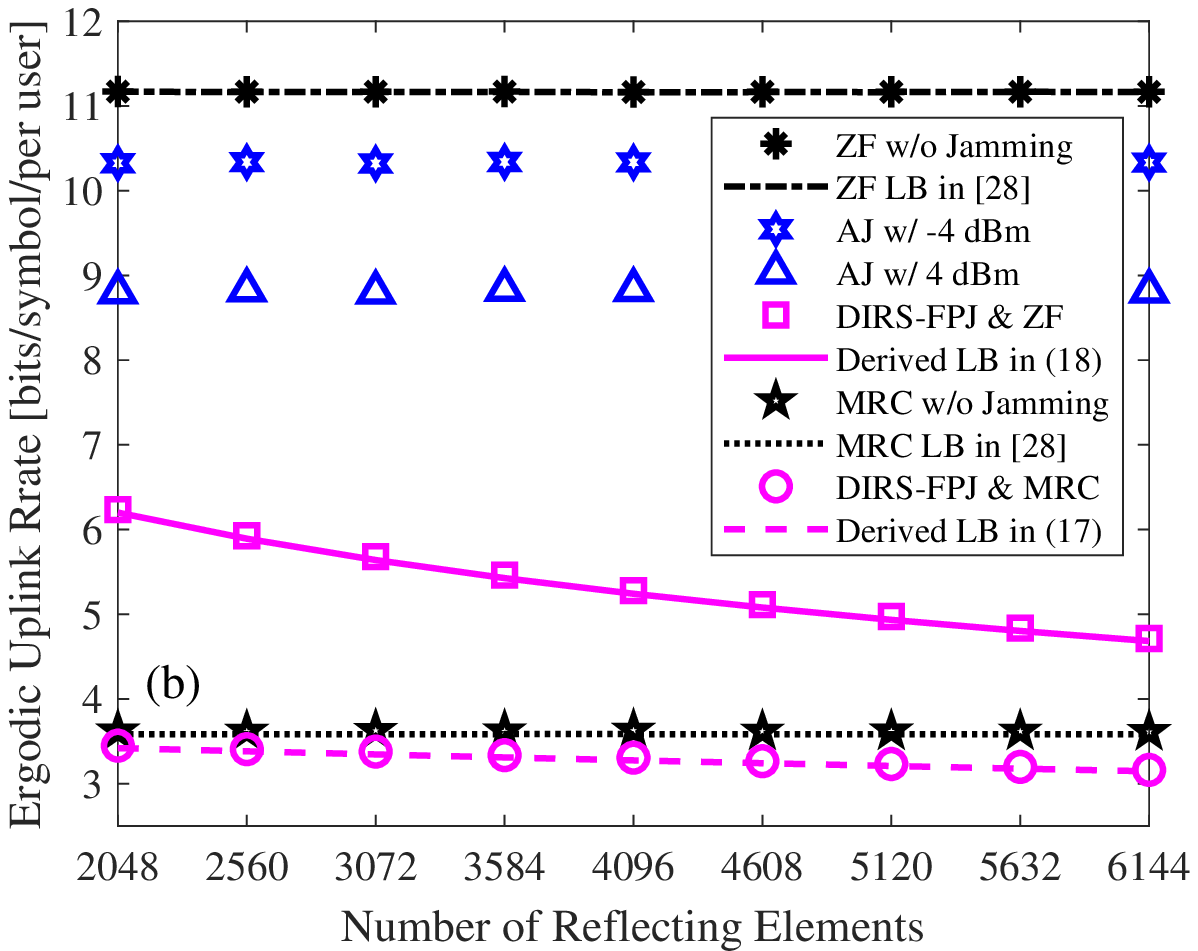}
     \label{fig5b}
 \end{minipage}
\caption{Influence of the number of DIRS reflecting elements on the ergodic rates for (a) -14 dBm average transmit power and (b) 6 dBm average transmit power.}
	\label{fig5}
\end{figure*}

Based on the theoretical analysis in Section~\ref{LowerBound}, the ergodic rates resulting from the MU-MISO system jammed by the proposed DIRS-based FPJ will tend to zero, as long as the number of DIRS reflecting elements is large. 
We show the effects of the number of DIRS reflecting elements in Fig.~\ref{fig5} at both low transmit power ($p_{\rm d} = -14$ dBm) and high transmit power ($p_{\rm d} = 6$ dBm), which are plotted in Fig.~\ref{fig5} (a) and Fig.~\ref{fig5} (b), respectively.

One can see that the MU-MISO system is sensitive to active jamming attacks when the average transmit power of each LU is low. However, the active jamming attacks can be suppressed by classic anti-jamming techniques such as frequency hopping~\cite{Frequencyhopping,Frequencyhopping1}, and an AJ requires a significant amount of extra jamming power.
The jamming impact is more effective for high rather than low transmit power.
\subsubsection{Ergodic Achievable Uplink Rate Versus DIRS Phase Resolution}
\begin{figure*}[!t]
\centering
 \begin{minipage}{0.65\linewidth}
     \centering
     \includegraphics[width=0.88\linewidth]{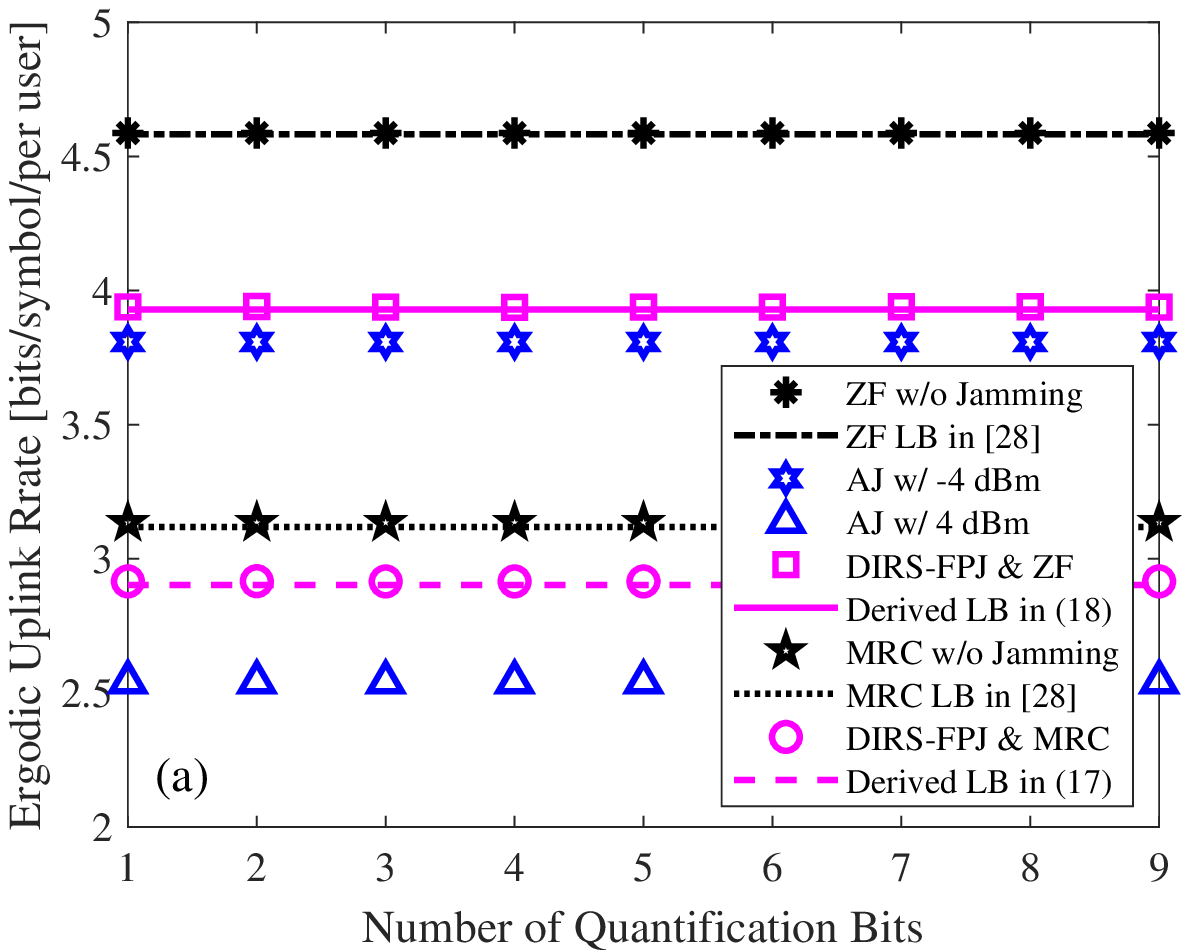}
     \label{fig7a}
 \end{minipage}

    \begin{minipage}{0.65\linewidth}
     \centering
     \includegraphics[width=0.88\linewidth]{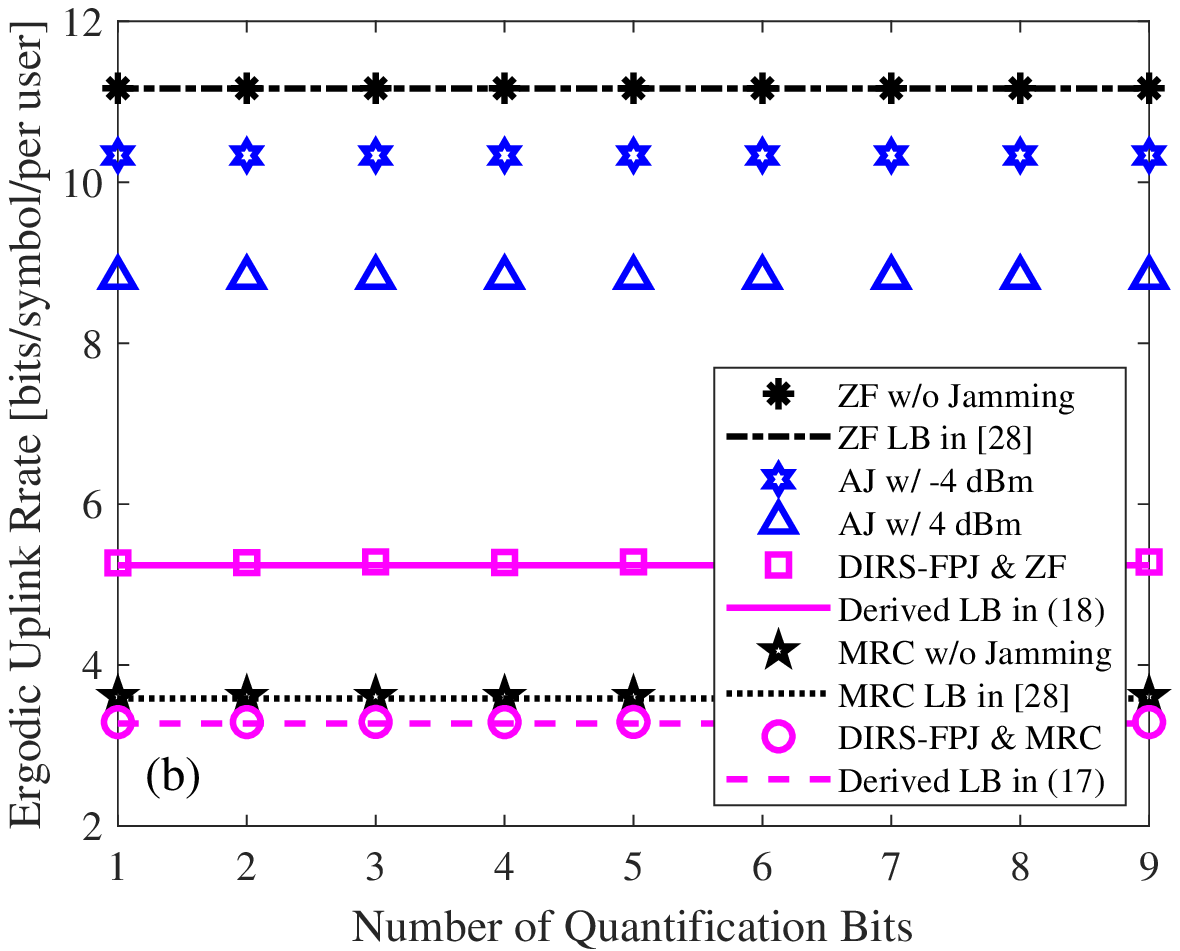}
     \label{fig7b}
 \end{minipage}
\caption{Influence of the number of DIRS phase quantification bits on the ergodic rates for (a) -14 dBm average transmit power and (b) 6 dBm average transmit power.}
	\label{fig7}
\end{figure*}

Based on the theoretical analysis in Section~\ref{LowerBound}, one can see that the proposed DIRS-based FPJ has many unique properties. For example, the jamming impact of the proposed DIRS-based FPJ does not depend on the resolution of the DIRS phase shifts. It is clear from Fig.~\ref{fig7} that the jamming impact of the proposed DIRS-based FPJ does not increase with the number of bits used to quantize the DIRS phase shifts for either low or high average transmit power. Based on the results in Fig.~\ref{fig5} and~\ref{fig7}, the proposed DIRS-based FPJ can be effectively implemented by using only a one-bit IRS with a large number reflecting elements whose phase randomly toggles between two values $\pi$ radians apart.
\subsubsection{Ergodic Achievable Uplink Rate Versus DIRS Location}
\begin{figure*}[!t]
\centering
 \begin{minipage}{0.65\linewidth}
     \centering
     \includegraphics[width=0.88\linewidth]{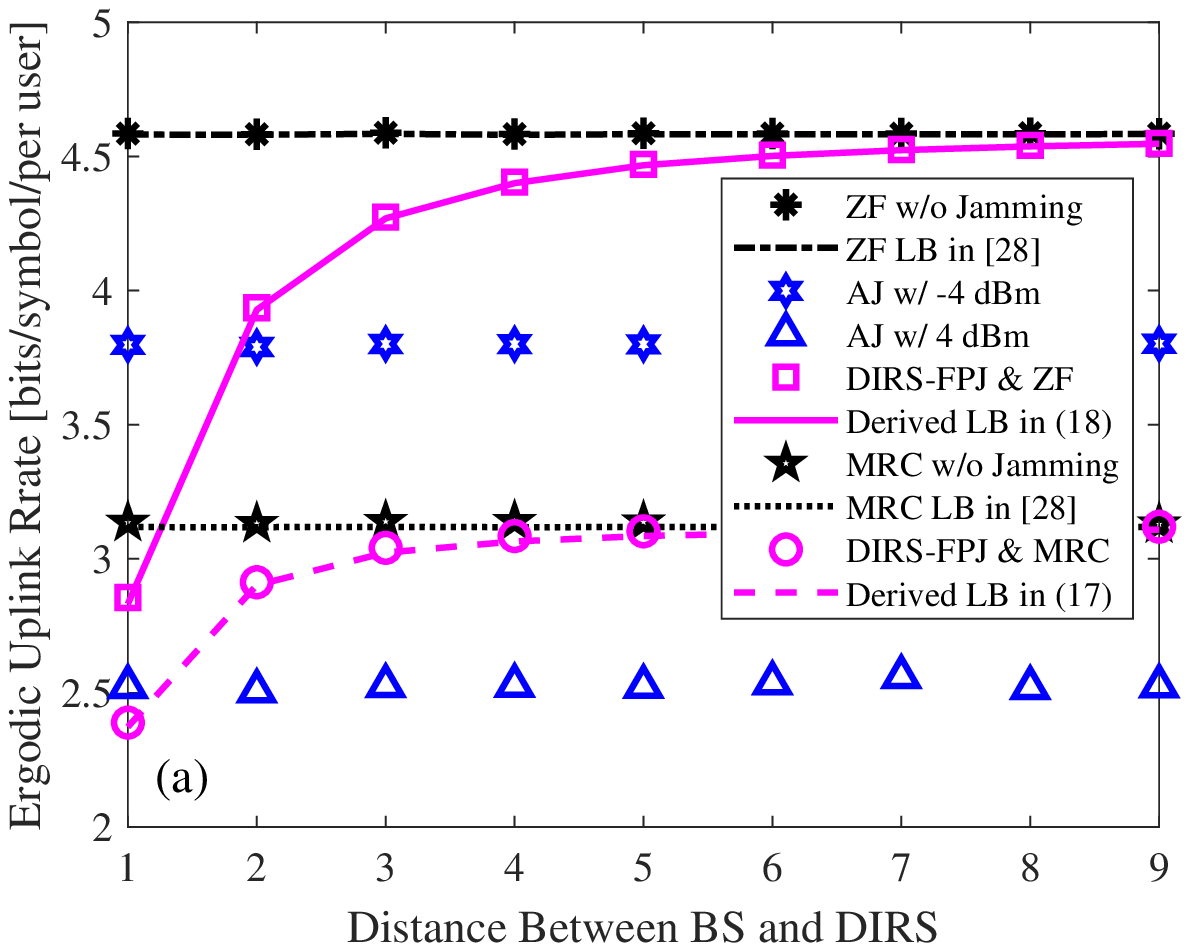}
     \label{fig7adda}
 \end{minipage}

    \begin{minipage}{0.65\linewidth}
     \centering
     \includegraphics[width=0.88\linewidth]{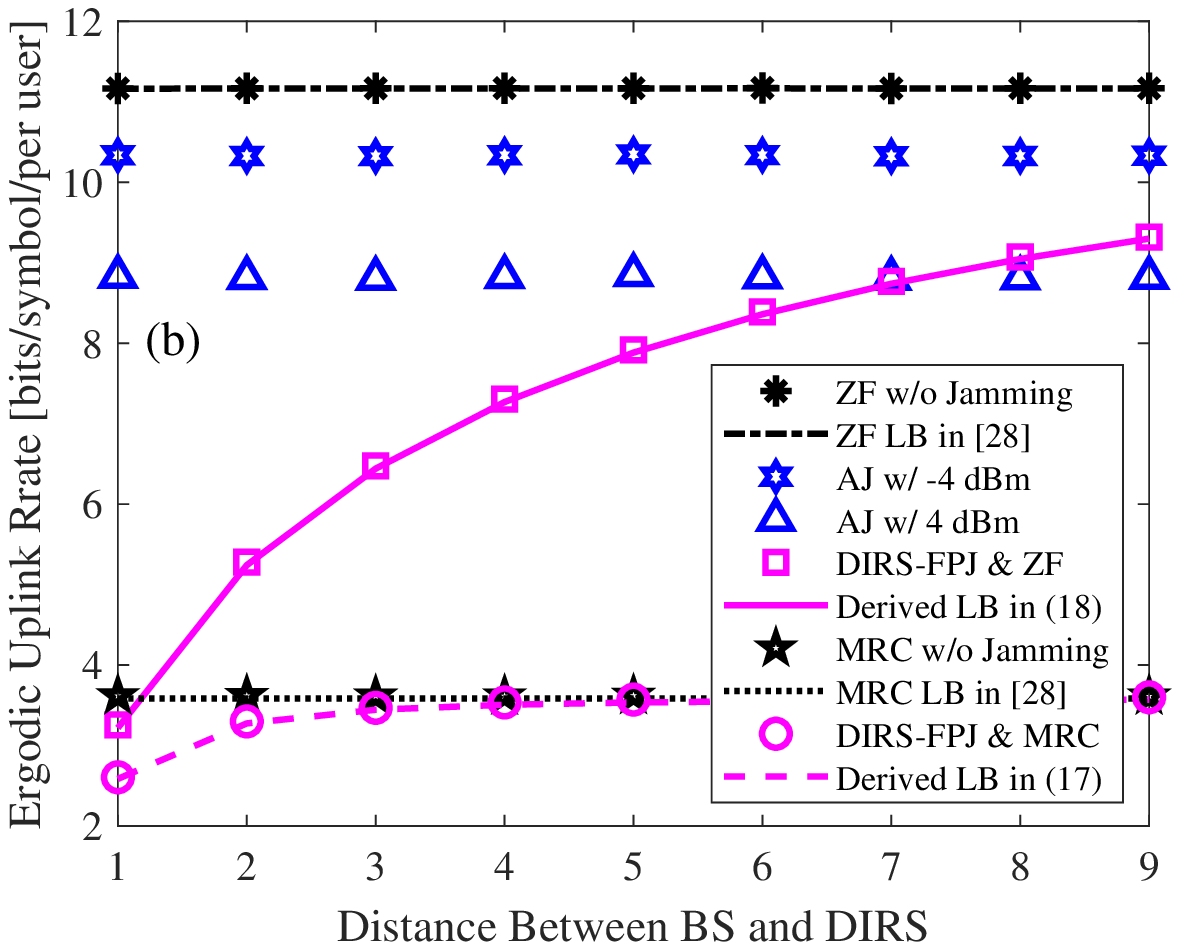}
     \label{fig7addb}
 \end{minipage}
\caption{Influence of the distance between the BS and the DIRS on the ergodic rates for (a) -14 dBm average transmit power and (b) 6 dBm average transmit power.}
	\label{fig7add}
\end{figure*}

In Fig.~\ref{fig7add}, the impact of the DIRS location on the ergodic rates is illustrated. The greater the distance, the greater the large-scale
channel fading ${\mathscr{L}}_{\rm G}$ in the BS-DIRS channel. According to \eqref{MRCLBeq} and \eqref{ZFLBeq}, the jamming effect of the proposed DIRS-based FPJ is weakened due to increased BS-DIRS distance $d_{\rm{DB}}$.
To maximize the jamming effect of the proposed DIRS-based FPJ, the DIRS needs to be deployed as close to the BS as possible. If a near-BS deployment is not possible, based on Fig.~\ref{fig5} and Fig.~\ref{fig7}, one solution to mitigating the weakening impact on jamming attacks is to increase the number of reflecting elements.
\subsubsection{Ergodic Achievable Uplink Rate Versus Number of BS Antennas}
\begin{figure*}[!t]
\centering
 \begin{minipage}{0.65\linewidth}
     \centering
     \includegraphics[width=0.88\linewidth]{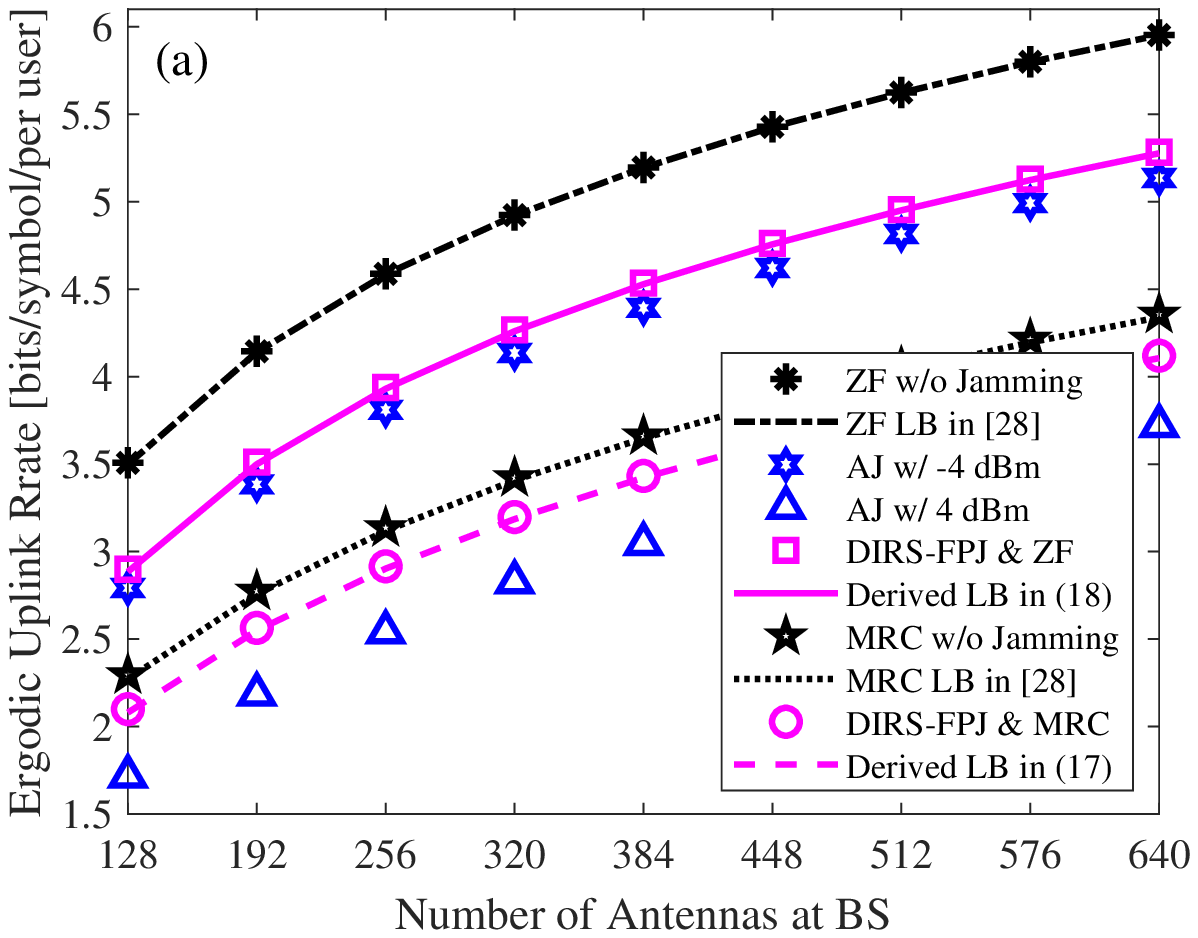}
     \label{fig6a}
 \end{minipage}

    \begin{minipage}{0.65\linewidth}
     \centering
     \includegraphics[width=0.88\linewidth]{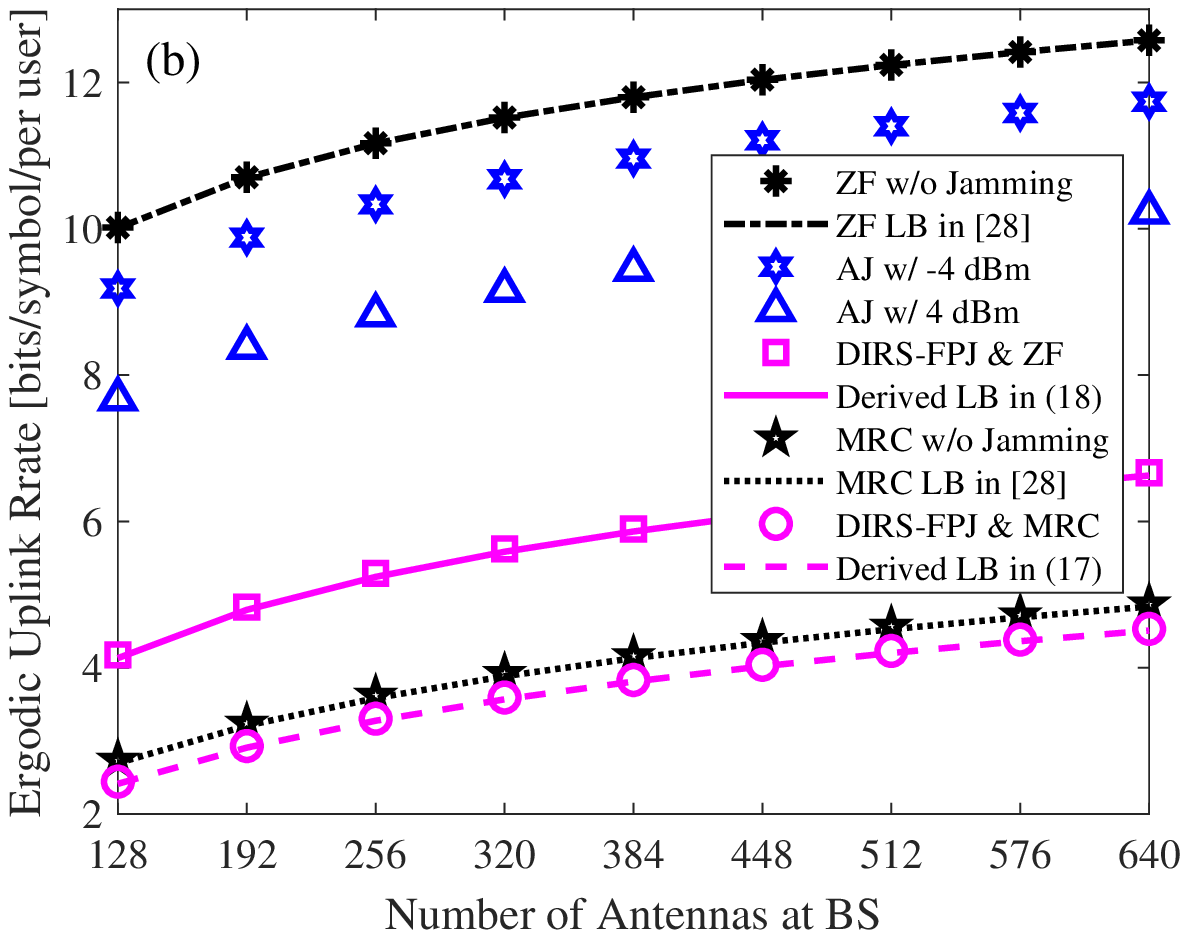}
     \label{fig6b}
 \end{minipage}
\caption{Influence of the number of antennas deployed at the BS on the ergodic rates for (a) -14 dBm average transmit power and (b) 6 dBm average transmit power.}
	\label{fig6}
\end{figure*}

In the next two figures, we demonstrate the jamming impact of the proposed DIRS-based FPJ on different MU-MISO systems. In Fig.~\ref{fig6}, we show the influence of the number of BS antennas on the ergodic rates. At both low and high transmit powers, the ergodic rates achieved by all benchmarks increase with the number of antennas at the BS. 

\begin{figure*}[!t]
\centering
 \begin{minipage}{0.65\linewidth}
     \centering
     \includegraphics[width=0.88\linewidth]{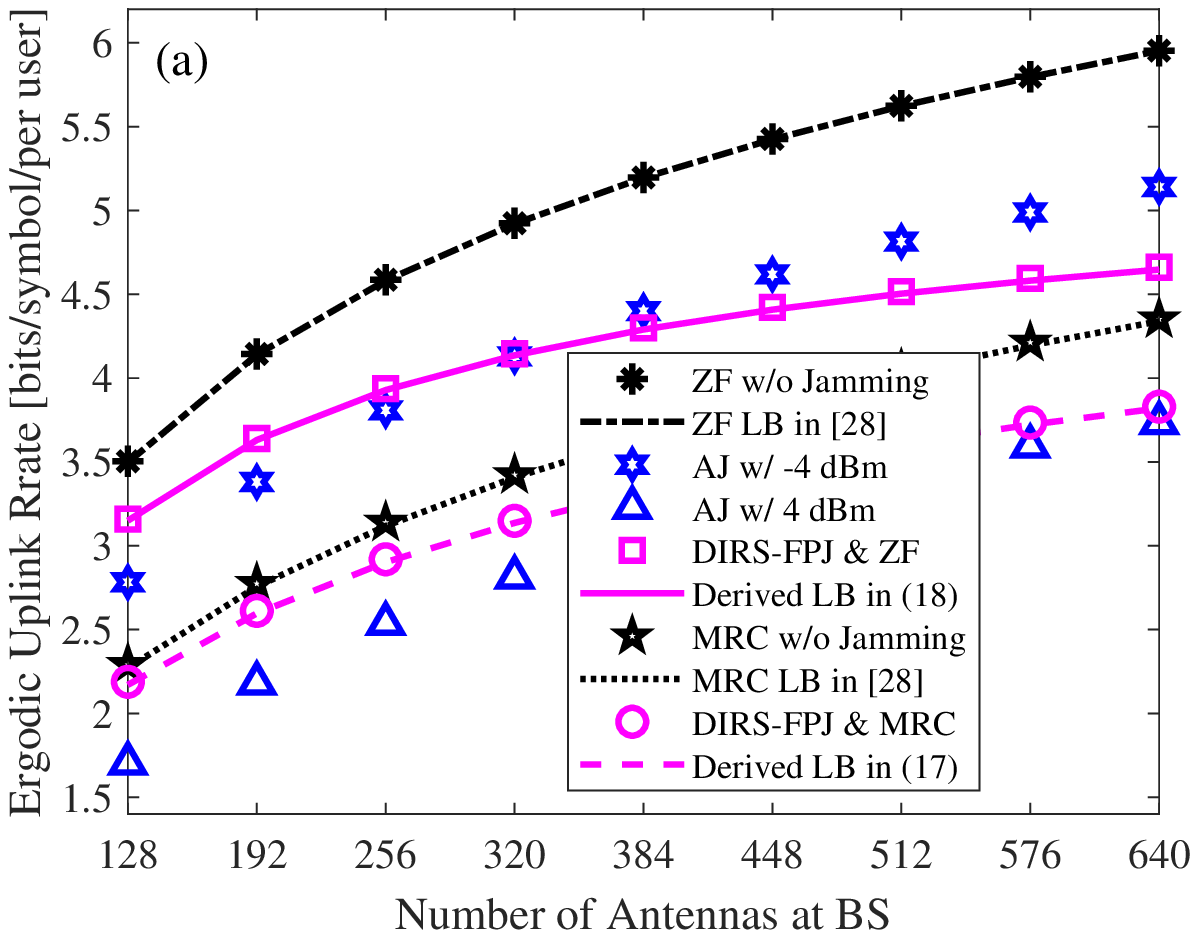}
     \label{fig61a}
 \end{minipage}

    \begin{minipage}{0.65\linewidth}
     \centering
     \includegraphics[width=0.88\linewidth]{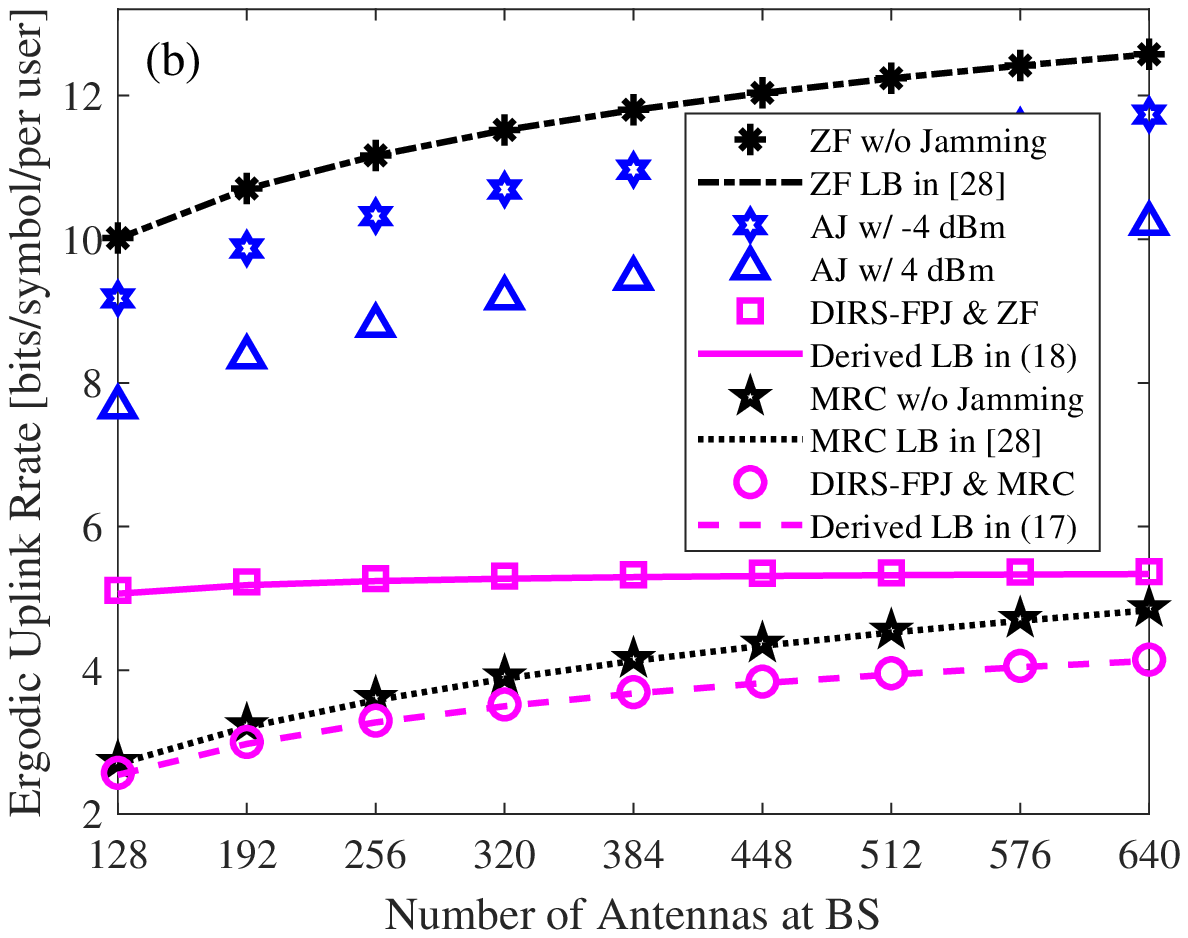}
     \label{fig61b}
 \end{minipage}
\caption{Ergodic rates versus numbers of antennas and reflecting elements for (a) -14 dBm average transmit power and (b) 6 dBm average transmit power, where the number of reflecting elements is sixteen times the number of antennas ($N_{\rm D} = 16N_t$).}
	\label{fig61}
\end{figure*}

A possible way to mitigate the jamming attacks launched by the proposed DIRS-based FPJ is to increase the number of BS antennas. However, we see in Figs.~\ref{fig6} and~\ref{fig61} that the slopes of all ergodic rate curves decrease as the number of BS antennas continues to increase. In other words, continuing to increase the number of antennas at the BS cannot significantly mitigate the proposed DIRS-based jamming attacks when the number of antennas is large. Moreover, we can increase the number of reflecting elements deployed on the DIRS to counteract this mitigation, as shown in Fig.~\ref{fig61}. It is clear that the ergodic rate resulting from the MU-MISO system jammed by the proposed DIRS-based FPJ does not increase with the number of BS antennas as long as the number of reflecting elements also increases.
Compared to the cost of increasing the number of active antennas at the BS, the cost of increasing the number of passive reflecting elements on the DIRS is much lower, especially for those employing one-bit phase shifters~\cite{RIS256ele,RISCB}.
\subsubsection{Ergodic Achievable Uplink Rate Versus Number of Legitimate Users}
\begin{figure*}[!t]
\centering
 \begin{minipage}{0.65\linewidth}
     \centering
     \includegraphics[width=0.88\linewidth]{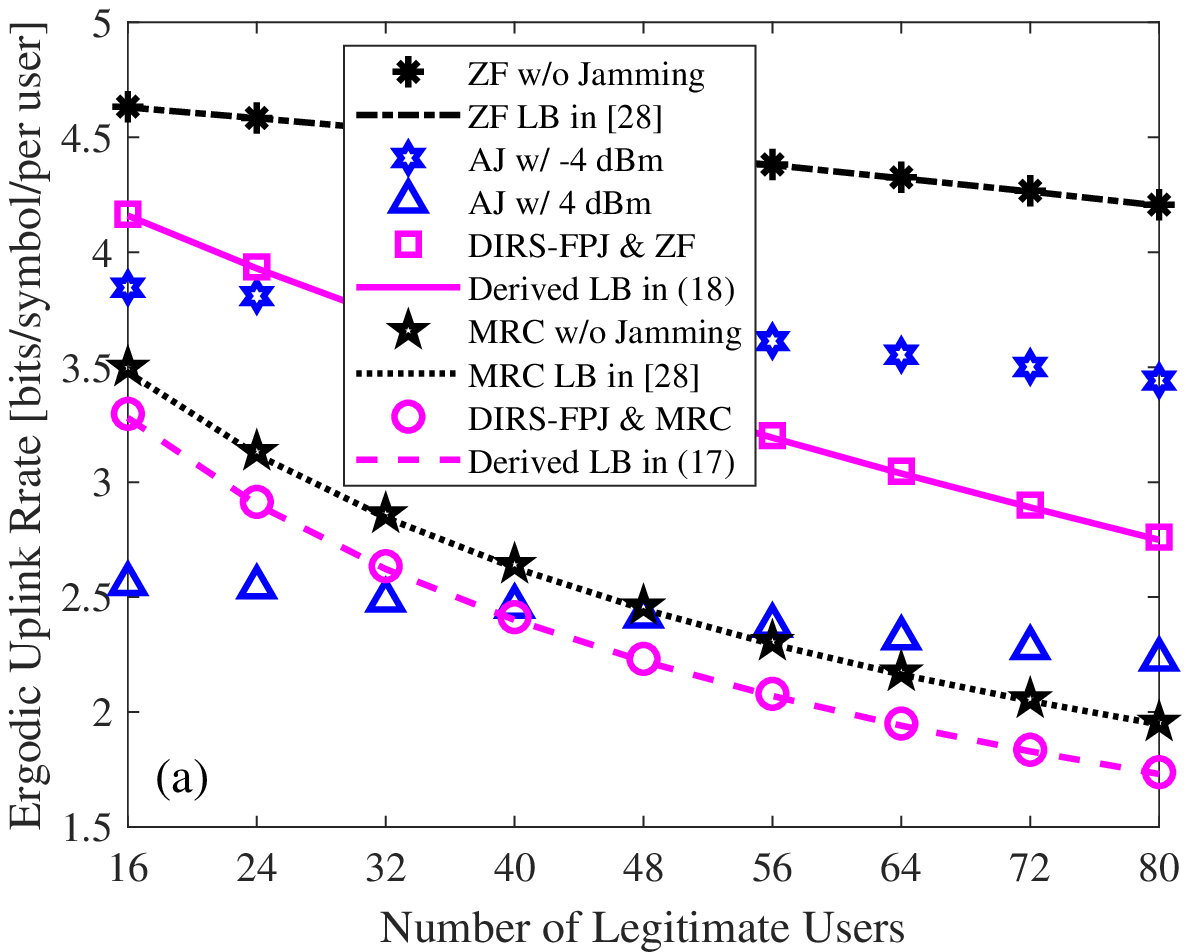}
     \label{fig8a}
 \end{minipage}

    \begin{minipage}{0.65\linewidth}
     \centering
     \includegraphics[width=0.88\linewidth]{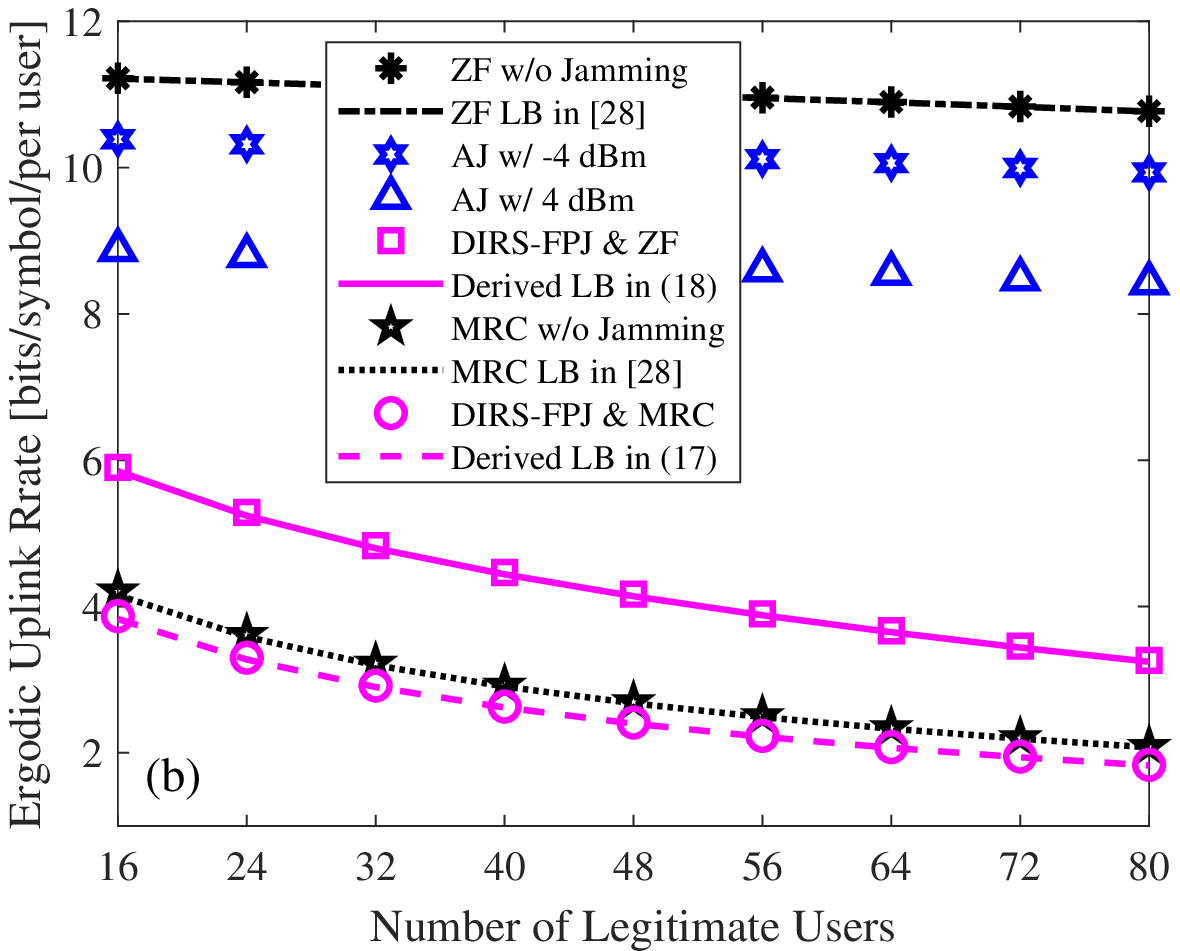}
     \label{fig8b}
 \end{minipage}
\caption{Influence of the number of LUs on the ergodic rates for (a) -14 dBm average transmit power and (b) 6 dBm average transmit power.}
	\label{fig8}
\end{figure*}

Fig.~\ref{fig8} shows the jamming impact of the proposed DIRS-based FPJ on MU-MISO systems communicating with different numbers of LUs. Even if an MU-MISO system is not subject to jamming attacks, the ergodic rate per LU decreases with the number of LUs. The drop in the ergodic rate is especially noticeable when the BS uses the MRC detector since it cannot suppress the IUI. 

At both low and high transmit power, the jamming impact of the proposed DIRS-based FPJ becomes more significant as the number of LUs increases. Note that the ACA interference term in~\eqref{kthsymeq} also increases with the number of LUs. Moreover, as the number of LUs increases, one can see that the gap between ZF w/o Jamming and DIRS-FPJ \& ZF is more significant than that between MRC w/o Jamming and DIRS-FPJ \& MRC at both low and high transmit power. As before, the increased IUI diminishes the impact of the ACA interference.
\section{Conclusions}\label{Conclu}
In this paper, a novel DIRS-based FPJ has been proposed that can be implemented by a one-bit IRS. By introducing significant ACA interference, the proposed DIRS-based FPJ can launch jamming attacks on LUs with neither extra jamming power nor LU CSI.
The following conclusions can be drawn from the theoretical analysis and numerical results, raising concerns about the significant potential threats posed by illegitimate IRSs.
\begin{enumerate}
\item  In contrast to existing AJs and IRS-based PJ, the proposed DIRS-based FPJ launches jamming attacks by using ACA interference caused by the DIRS, and thus it requires no jamming power and no LU CSI.

\item The jamming impact of the proposed DIRS-based FPJ does not depend on how the DIRS phase shifts are quantized, nor on their distribution. As long as the number of DIRS reflecting elements is large, the ergodic rate will tend to zero even if the proposed DIRS-based FPJ is implemented with one-bit uniformly distributed phase shifts.

\item Increasing the transmit power at each LU will not mitigate the jamming attacks launched by the proposed DIRS-based FPJ and will even make them more deleterious. Furthermore, the  DIRS-based FPJ can overcome classic anti-jamming technologies such as frequency hopping.

\item Although the BS can counter the proposed DIRS-based jamming attacks by increasing the number of its antennas, this suppression of the proposed DIRS-based jamming attacks can be weakened by increasing the number of reflecting elements on the DIRS.
\end{enumerate}
We have illustrated the potential threats posed by illegal IRSs and demonstrated that even a one-bit DIRS can effectively jam LUs. Classic anti-jamming technologies are not effective for the proposed DIRS-based FPJ. 

\appendices
\section{Proof of Proposition~\ref{Proposition2} }\label{AppendixA}
Recall the ergodic ${\left. {{{\overline R }_{{\rm{d}},k}}} \right|_{{\rm{MRC}}}}$ expressed as~\eqref{GrgGamakMRCeq}. We have the following lower bound by using Jensen's inequality:
\begin{alignat}{1}
\nonumber
{\left. {{{\overline R }_{{\rm{d}},k}}} \right|_{{\rm{MRC}}}} & = \mathbb{E}\left[ {{{\log }_2}\left( {1  +  {\left. {{\gamma_k}} \right|_{{\rm{MRC}}}}} \right)} \right] \\
\nonumber
&\ge {{{\log }_2}\left( {1 + \frac{1}{{\mathbb{E}\left[ {{{\left( {{{\left. {{\gamma_k}} \right|}_{{\rm{MRC}}}}} \right)}^{ - 1}}} \right]}}} \right)} \\
&= {\log _2}\left( {1 + \frac{1}{{\mathbb{E}\!\!\left[\! {\frac{{{p_{\rm{d}}}\sum\nolimits_{i = 1,i \ne k}^K {{{ \left|{{{\widetilde h}_{{\rm{d}},i}}}\right| }^2}}  + {p_{\rm{d}}}\sum\nolimits_{i = 1}^K {{{ \left|{{{\widetilde h}_{{\rm{D}},i}}}\right| }^2}} + \sigma _{\rm{d}}^2}}{{{p_{\rm{d}}}{{\left\| {{\boldsymbol{h}_{{\rm{d}},k}}} \right\|}^2}}}} \right]}}} \right),
\label{JensenMRCeq}
\end{alignat}
where ${{{\widetilde h}_{{\rm{d}},i}}}  = {\frac{{{{ {{\boldsymbol{h}}_{{\rm{d}},k}^H{{\boldsymbol{h}}_{{\rm{d}},i}}} } }}}{{{{\left\| {{{\boldsymbol{h}}_{{\rm{d}},k}}} \right\|} }}}}$ and ${{{\widetilde h}_{{\rm{D}},i}}} = {\frac{{{{ {{\boldsymbol{h}}_{{\rm{d}},k}^H{{\boldsymbol{h}}_{{\rm{D}},i}}} }}}}{{{{\left\| {{{\boldsymbol{h}}_{{\rm{d}},k}}} \right\|}}}}}$.
Conditioned on the fact that the random variables ${\left\| {{\boldsymbol{h}_{{\rm{d}},k}}} \right\|}$, ${{{\widetilde h}_{{\rm{d}},i}}}$, and ${{{\widetilde h}_{{\rm{D}},i}}}$ are independent, we can reduce the expectation in~\eqref{JensenMRCeq} to
\begin{alignat}{1}
\nonumber
&{{\mathbb{E}\!\!\left[\! {\frac{{{p_{\rm{d}}}\!\!\sum\limits_{i = 1,i \ne k}^K {{{ \!\left|{{{\widetilde h}_{{\rm{d}},i}}}\right| }^2}}\!  + {p_{\rm{d}}}\sum\limits_{i = 1}^K {{{ \!\left|{{{\widetilde h}_{{\rm{D}},i}}}\right| }^2}} + \sigma _{\rm{d}}^2}}{{{p_{\rm{d}}}{{\left\| {{\boldsymbol{h}_{{\rm{d}},k}}} \right\|}^2}}}} \right]}} \\
& \;\;\;\;\;\;\;\;\;\;\;\;\;\;\;\;\; = \left({{p_{\rm{d}}}\sum\limits_{i = 1,i \ne k}^K {  \mathbb{E}\!\left[{{\left| {{{\widetilde h}_{{\rm{d}},i}}}\right| }^2}\right] }  + {p_{\rm{d}}}\sum\limits_{i = 1}^K {  \mathbb{E}\!\left[{{ \left|{{{\widetilde h}_{{\rm{D}},i}}}\right| }^2}\right] } + \sigma _{\rm{d}}^2}\right) {{\mathbb{E}\!\!\left[\! {\frac{1}{{{p_{\rm{d}}}{{\left\| {{\boldsymbol{h}_{{\rm{d}},k}}} \right\|}^2}}}} \right]}} \label{indepenteqadd}.
\end{alignat}

Based on the weak law of large numbers, the random vector $\frac{{{{\boldsymbol{h}}_{{\rm{d}},i}}}}{{\left\| {{{\boldsymbol{h}}_{{\rm{d}},k}}} \right\|}}$ with i.i.d. elements converges in probability towards $\frac{{{{\boldsymbol{h}}_{{\rm{d}},i}}}} {\sqrt{ {\mathscr{L}_{{\rm{d}},k}} N_t}}$ as $N_t \to \infty$, i.e.,
\begin{equation}
\frac{{{{\boldsymbol{h}}_{{\rm{d}},i}}}}{{\left\| {{{\boldsymbol{h}}_{{\rm{d}},k}}} \right\|}} = \frac{{{{\boldsymbol{h}}_{{\rm{d}},i}}}}{\sqrt {{\mathscr{L}_{{\rm{d}},k}}\sum\limits_{n = 1}^{{N_t}} \left|\!{\left[ {{{\widehat {\bf{H}}}_{\rm{d}}}} \right]_{nk}}\right|^2 }}  \mathop  \to \limits^{\rm{p}} \frac{{{{\boldsymbol{h}}_{{\rm{d}},i}}}} {\sqrt{{\mathscr{L}_{{\rm{d}},k}}N_t }}, \;{\rm{as}}\;N_{t} \to \infty.
\label{LLNnormdeq}
\end{equation}
According the central limit theorem, the random variable ${\frac{{{{ {{\boldsymbol{h}}_{{\rm{d}},k}^H{{\boldsymbol{h}}_{{\rm{d}},i}}} } }}}{ \sqrt{N_t} }}$ converges in distribution to a normal $\mathcal{CN}\left( {0, {{{\mathscr{L}_{{\rm{d}},k}}{\mathscr{L}_{{\rm{d}},i}}}} } \right)$ as $N_t \to \infty$, i.e.,
\begin{equation}
{\frac{{{{{{\boldsymbol{h}}_{{\rm{d}},k}^H{{\boldsymbol{h}}_{{\rm{d}},i}}}} }}}{ \sqrt{N_t} }} \mathop  \to \limits^{\rm{d}} \mathcal{CN}\left( {0, {{{\mathscr{L}_{{\rm{d}},k}}{\mathscr{L}_{{\rm{d}},i}}}} } \right),\;{\rm{as}}\; N_t \to \infty.
\label{CLTLLNeq2}
\end{equation}
Based on~\eqref{LLNnormdeq} and~\eqref{CLTLLNeq2}, the random variable ${{{\widetilde h}_{{\rm{d}},i}}}$ converges in distribution to $\mathcal{CN}\left( {0, {\mathscr{L}_{{\rm{d}},i}} } \right)$ as $N_t \to \infty$,
\begin{equation}
{{{\widetilde h}_{{\rm{d}},i}}} \mathop  \to \limits^{\rm{d}} \mathcal{CN}\left( {0, {{\mathscr{L}_{{\rm{d}},i}}} } \right),\;{\rm{as}}\; N_t \to \infty \label{CLTLLNeq11} .
\end{equation}
Consequently, the term ${p_{\rm{d}}}\sum\nolimits_{i = 1,i \ne k}^K { \mathbb{E}\!\left[{{\left| {{{\widetilde h}_{{\rm{d}},i}}}\right| }^2}\right] }$ in~\eqref{indepenteqadd} is reduced to ${p_{\rm{d}}}\sum\nolimits_{i = 1,i \ne k}^K {{\mathscr{L}_{{\rm{d}},i}}}$.

On the other hand, from~\eqref{Ricianchan} to~\eqref{HIkeq}, the i.i.d. elements in the multi-user DIRS-based channel ${\bf{H}}_{\rm D}$ can be written as
\begin{equation}
{\left[ {{{\bf{H}}_{\rm{D}}}} \right]_{nk}} = \sqrt {\frac{{{\varepsilon_n}{\mathscr{L}_{\rm{G}}}{\mathscr{L}_{{\rm{I}},k}}}}{{{\varepsilon_n} + 1}}} \sum\limits_{r = 1}^{{N_{\rm{D}}}} {{e^{ - j\frac{{2\pi }}{\lambda }d\left( {n - 1} \right)\sin {\theta _r}}}{e^{j{\varphi _r}}}{{\left[ {{{\widehat {\bf{H}}}_{\rm{I}}}} \right]}_{rk}}} + \sqrt {\frac{{{\mathscr{L}_{\rm{G}}}{\mathscr{L}_{{\rm{I}},k}}}}{{{\varepsilon_n} + 1}}} \sum\limits_{r = 1}^{{N_{\rm{D}}}} {\left[ {{{\bf{G}}^{{\rm{NLOS}}}}} \right]_{rn}^H{e^{j{\varphi _r}}}{{\left[ {{{\widehat {\bf{H}}}_{\rm{I}}}} \right]}_{rk}}}.
\label{HDelements}
\end{equation}
Assume that the elements in ${\bf{H}}_{\rm D}$ and ${\bf{H}}_{\rm d}$ are independent; therefore, we have that
\begin{equation}
\mathbb{E}\!\left[\left|{{{{{\boldsymbol{h}}_{{\rm{d}},i}^H{{\boldsymbol{h}}_{{\rm{D}},k}}} } }}\right|^2\right] =  \mathbb{E}\!\left[{\sum\limits_{n = 1}^{{N_t}} {\mathscr{L}_{{\rm{d}},i}}{\left(\left| { {\left[ {{{\widehat {\bf{H}}}_{\rm{d}}}} \right]}_{ni}}  {{{\left[ {{{\bf{H}}_{\rm{D}}}} \right]}_{nk}} }\right|^2 \right)} }\right] =  {\mathscr{L}_{{\rm{d}},i}} \sum\limits_{n = 1}^{{N_t}} \mathbb{E}\!\left[{ \left|  {{{\left[ {{{\bf{H}}_{\rm{D}}}} \right]}_{nk}} }\right|^2  }\right] .
\label{hDhdexpeq}
\end{equation}
According to~\eqref{HDelements}, the expectation of ${ \left|  {{{\left[ {{{\bf{H}}_{\rm{D}}}} \right]}_{nk}} }\right|^2  }$ in~\eqref{hDhdexpeq} is given by,
\begin{equation}
\mathbb{E}\!\left[ {{{\left| {{{\left[ {{{\bf{H}}_{\rm{D}}}} \right]}_{nk}}} \right|}^2}} \right] = \mathbb{E}\!\left[ {\left[ {{{\bf{H}}_{\rm{D}}}} \right]_{nk}^H{{\left[ {{{\bf{H}}_{\rm{D}}}} \right]}_{nk}}} \right] = {\mathscr{L}_{\rm{G}}}{\mathscr{L}_{{\rm{I}},k}}{N_{\rm{D}}}.
\label{expelementHD}
\end{equation}
As a result, the expectation ${p_{\rm{d}}}\sum\nolimits_{i = 1}^K {{{ \left|{{{\widetilde h}_{{\rm{D}},i}}}\right| }^2}} $ in~\eqref{indepenteqadd} is reduced to ${p_{\rm{d}}}{N_{\rm D}}\sum\nolimits_{i = 1}^K {{{ {{\mathscr{L}_{\rm{G}}}{\mathscr{L}_{{\rm{I}},i}}} } }}$. Furthermore, we can reduce the expectation in~\eqref{JensenMRCeq} to
\begin{alignat}{1}
& \mathbb{E}\!\left[ {{p_{\rm{d}}}\sum\limits_{i = 1,i \ne k}^K {{{\left| {{{\widetilde h}_{{\rm{d}},i}}}\right| }^2}}  + {p_{\rm{d}}}\sum\limits_{i = 1}^K {{{ \left|{{{\widetilde h}_{{\rm{D}},i}}}\right| }^2}} + \sigma _{\rm{d}}^2} \right] {{\mathbb{E}\!\!\left[\! {\frac{1}{{{p_{\rm{d}}}{{\left\| {{\boldsymbol{h}_{{\rm{d}},k}}} \right\|}^2}}}} \right]}} \label{indepenteq}\\
&\;\;\;\;\;\;\;\;\;\;\;\;\;\;\;\;\;\;\; \mathop  \to \limits^{\rm{p}} \!\left(\!{{{p_{\rm{d}}} \!\sum\limits_{i = 1,i \ne k}^K \!{\!{\mathscr{L}_{{\rm{d}},i}}}  + {p_{\rm{d}}}{N_{\rm{D}}}\sum\limits_{i = 1}^K \!{{\mathscr{L}_{\rm{G}}}{\mathscr{L}_{{\rm{I}},i}}}  + \sigma _{\rm{d}}^2}} \right) \!{\mathbb{E}\!\!\left[\! {\frac{1}{{{p_{\rm{d}}}{{\left\| {{\boldsymbol{h}_{{\rm{d}},k}}} \right\|}^2}}}} \right]} \label{indepenteq1}.
\end{alignat}

Next we exploit the following property of a central complex Wishart matrix~\cite{Wishart}, i.e.,
\begin{equation}
\mathbb{E}\left[ {{\rm{tr}}\left( {{{\bf{W}}^{ - 1}}} \right)} \right] = \frac{m}{{n - m}},
\label{Wisharteq}
\end{equation}
where ${\bf{W}} \sim {\cal W}\left( {n,{{\bf{I}}_n}} \right)$ is an $m \times m$ central complex Wishart matrix with $n$ ($n > m$) degrees of freedom.
Incorporating~\eqref{Wisharteq} into~\eqref{indepenteq1}, the following equation is obtained:
\begin{equation}
{\mathbb{E}\!\!\left[\! {\frac{1}{{{p_{\rm{d}}}{{\left\| {{\boldsymbol{h}_{{\rm{d}},k}}} \right\|}^2}}}} \right]} = \frac{1}{{{p_{\rm{d}}}\left( {{N_t} - 1} \right){\mathscr{L}_{{\rm{d}},k}}}},\; {\rm {for}} N_t > 2.
\label{ExpWisharteq}
\end{equation}
Combining~\eqref{ExpWisharteq} and~\eqref{indepenteq1}, the lower bound in Proposition~\ref{Proposition2}, i.e., \eqref{MRCLBeq} is then derived.
\section{Proof of Proposition~\ref{Proposition3} }\label{AppendixB}
Based on Jensen's inequality, the ergodic rate ${\left. {{{\overline R }_{{\rm{d}},k}}} \right|_{{\rm{ZF}}}}$ in~\eqref{GrgGamakZFeq} reduces to
\begin{equation}
{\left. {{{\overline R }_{{\rm{d}},k}}} \right|_{{\rm{ZF}}}}  \ge {{{\log }_2}\!\left(\! {1 + \frac{{{p_{\rm{d}}}}}{\mathbb{E}\!\left[{{p_{\rm{d}}} \sum\limits_{i = 1}^K {{\left| {{\boldsymbol{w}_k^H}{\boldsymbol{h}_{{\rm{D}},i}}} \right|}^2} }\right] + \mathbb{E}\!\left[{{\sigma_{\rm d}^2}{\left\| {{\boldsymbol{w}_k}} \right\|^2}} \right]}} \right)} .
\label{JenseGamakZFeq}
\end{equation}
We can assume that the random vector $\boldsymbol{w}_k$ does not depend on ${\boldsymbol{h}_{{\rm{D}},i}}$, because the linear detector ZF is designed based on the multi-user direct channel ${\bf H}_{\rm d}$ without relying on ${\bf{H}_{{\rm{D}}}}$.
Therefore, based on the definition in~\eqref{HDelements}, we have the following result:
\begin{equation}
\mathbb{E}\!\left[{{p_{\rm{d}}} \sum\limits_{i = 1}^K {{\left| {{\boldsymbol{w}_k^H}{\boldsymbol{h}_{{\rm{D}},i}}} \right|}^2} }\right] = {p_{\rm{d}}}\sum\limits_{i = 1}^K {\sum\limits_{n = 1}^{{N_t}} {\mathbb{E}\!\left[ {{{\left| {{w_{kn}}} \right|}^2}} \right]} } \mathbb{E}\!\left[ {{{\left| {{{\left[ {{{\bf{H}}_{\rm{D}}}} \right]}_{ni}}} \right|}^2}} \right],
\label{JenseGamakZFeleeq}
\end{equation}
where ${w_{kn}}$ denotes the $n$-th element of the ZF detector vector ${\boldsymbol{w}_k}$.

According to~\eqref{expelementHD} and~\eqref{JenseGamakZFeleeq}, \eqref{JenseGamakZFeq} is then converted to
\begin{alignat}{1}
\nonumber
{\left. {{{\overline R }_{{\rm{d}},k}}} \right|_{{\rm{ZF}}}}  &\ge {{{\log }_2}\!\left(\! {1 + \frac{{{p_{\rm{d}}}}}{\mathbb{E}\left[{{p_{\rm{d}}} \sum\limits_{i = 1}^K {{\left| {{\boldsymbol{w}_k^H}{\boldsymbol{h}_{{\rm{D}},i}}} \right|}^2} }\right] + \mathbb{E}\left[{{\sigma_{\rm d}^2}{\left\| {{\boldsymbol{w}_k}} \right\|^2}} \right]}} \right)}  \\
&  = {{{\log }_2}\!\left(\! {1 + \frac{{{p_{\rm{d}}}}}{ \left({p_{\rm{d}}}{N_{\rm{D}}} \!\sum\limits_{i = 1}^K \!{{\mathscr{L}_{{\rm{G}}}}{\mathscr{L}_{{\rm{I}},i}}} + {\sigma_{\rm d}^2}\right)   \mathbb{E}\!\left[{{{\left\|{{\boldsymbol{w}_k}} \right\|^2}} }\right] }} \right)} \label{JenseGamakZFeq1}.
\end{alignat}
Based on~\eqref{ZFeq}, the following form of ${\left\| {{\boldsymbol{w}_k}} \right\|}$  can be obtained:
\begin{equation}
{\left\| {{\boldsymbol{w}_k}} \right\|^2} = {\left[ {{{\bf{W}}^H}{\bf{W}}} \right]_{kk}} = \frac{1}{{{\mathscr{L}_{{\rm{d}},k}}}}{\left[ {{{\left( {\widehat {\bf{H}}_{\rm{d}}^H{{\widehat {\bf{H}}}_{\rm{d}}}} \right)}^{ - 1}}} \right]_{kk}}.
\label{effZFeq}
\end{equation}
Consequently,
\begin{alignat}{1}
\mathbb{E}\left[ {{{\left\| {{\boldsymbol{w}_k}} \right\|}^2}} \right] &= \frac{1}{{K{\mathscr{L}_{{\rm{d}},k}}}}\mathbb{E}\left[ {{\rm{tr}}{{\left( {\widehat {\bf{H}}_{\rm{d}}^H{{\widehat {\bf{H}}}_{\rm{d}}}} \right)}^{ - 1}}} \right] \label{ZFqweq}  \\
&\mathop  = \limits^{(a)} \frac{1}{{\left( {{N_t} - K} \right){\mathscr{L}_{{\rm{d}},k}}}}, \label{ExprqiZFeq1}
\end{alignat}
where $(a)$ also comes from the identity~\eqref{Wisharteq}.

Substituting~\eqref{ExprqiZFeq1} into~\eqref{JenseGamakZFeq1}, the lower bound~\eqref{ZFLBeq} based on the use of ZF is derived.


\begin{thebibliography}{1}
\bibliographystyle{IEEEtran}
\bibitem{PLSsur1}
A.~Mukherjee, S.~A.~A.~Fakoorian, J.~Huang, and A.~L.~Swindlehurst, ``Principles of physical layer security in multiuser wireless networks: A survey," \emph{IEEE Commun. Surv. Tut.}, vol. 16, no. 3, pp. 1550--1573, Third Quarter 2014.

\bibitem{DoSsur1}
Y.~Zou, J.~Zhu, X.~Wang, and L.~Hanzo, ``A survey on wireless security: Technical challenges, recent advances, and future trends," \emph{Proceedings of the IEEE}, vol. 104, no. 9, pp. 1727--1765, Spet. 2016.

\bibitem{AntiJammingSurv}
H.~Pirayesh and H.~Zeng, ``Jamming attacks and anti-jamming strategies in wireless networks: A comprehensive survey," \emph{IEEE Commun. Surv. Tut.}, vol. 24, no. 2, pp. 767--809, 2nd Quarter 2022.

\bibitem{crypbook}
P.~Christof, J.~Pelzl, and B.~Prenee, \emph{Understanding Cryptography: A Textbook for Students and Practitioners}. New York, NY, USA: Springer-Verlag, 2010.

\bibitem{DoSJamm}
H.~Zhang, P.~Cheng, L.~Shi, and J.~Chen, ``Optimal denial-of-service attack scheduling with energy constraint," \emph{IEEE Trans. Autom. Control}, vol. 60, no. 11, pp. 3023--3028, Nov. 2015.

\bibitem{DoSJamm2}
H.~Zhang, P.~Cheng, L.~Shi, and J.~Chen, ``Optimal DoS attack scheduling in wireless networked control system," \emph{IEEE Trans. Control Syst. Technol.}, vol. 24, no. 3, pp. 843--852, May 2016.

\bibitem{intermittentAJ}
O.~Besson, P.~Stoica, and Y.~Kamiya, ``Direction finding in the presence of an intermittent interference," \emph{IEEE Trans. Signal Process.}, vol. 50, no. 7, pp. 1554--1564, Jul. 2002.

\bibitem{reactiveAJ}
E.~Lance and G.~K.~Kaleh, ``A diversity scheme for a phase-coherent frequency-hopping spread-spectrum system," \emph{IEEE Trans. Commun.}, vol. 45, no. 9, pp. 1123--1129, Sep. 1997.

\bibitem{adaptiveAJ}
J.~Jeung, S.~Jeong, and J.~Lim, ``Adaptive rapid channel-hopping scheme mitigating smart jammer attacks in secure WLAN," in \emph{Proc. Military Commun. Conf.}, Baltimore, MD, Nov. 2011, pp. 1231--1236.

\bibitem{IRSsuradd}
S.~Tomasin, H.~Zhang, A.~Chorti, and H.~V.~Poor, ``Challenge-response physical layer authentication over partially controllable channels," \emph{IEEE Commun. Mag.}, vol. 60, no. 12, pp. 138--144, Dec. 2022.

\bibitem{IRSsur1}
H.~Zhang and B.~Di, ``Intelligent omni-surfaces: Simultaneous refraction and reflection for full-dimensional wireless communication," \emph{IEEE Commun. Surv. Tut.}, vol. 24, no. 4, pp. 1997--2028, Aug. 2022.

\bibitem{IRSsur11}
Q.~Wu, S.~Zhang, B.~Zheng, C.~You, and R.~Zhang, ``Intelligent reflecting surface aided wireless communications: A tutorial," \emph{IEEE Trans. Commun.}, vol. 69, no. 5, pp. 3313--3351, Jan. 2021.

\bibitem{IRSsur2}
M.~A.~ElMossallamy, H.~Zhang, L.~Song, K.~G.~Seddik, Z.~Han, G.~Y.~Li, ``Reconfigurable intelligent surfaces for wireless communications: Principles, challenges, and opportunities," \emph{IEEE Trans. Cogn. Commun.}, vol. 6, no. 3, pp. 990--1002, Sept. 2020.

\bibitem{IRSsur3}
S.~Gong, X.~Lu, D. T.~Hoang, D.~Niyato, L.~Shu, D. I.~Kim, Y.-C.~Liang, ``Toward smart wireless communications via intelligent reflecting surfaces: A contemporary survey," \emph{IEEE Commun. Surv. Tut.}, vol. 22, no. 4, pp. 2283--2314, Fourth Quarter 2020.

\bibitem{IRSdeployment}
S.~Zhang and R.~Zhang, ``Intelligent reflecting surface aided multi-user comunication: Capacity region and deployment strategy," \emph{IEEE Trans. Commun.}, vol. 69, no. 9, pp. 5790--5806, Spet. 2021.

\bibitem{HuangRIS}
C.~Huang, S.~Hu, G.~C.~Alexandropoulos, A.~Zappone, C.~Yuen, R.~Zhang, M.~Di Renzo, and M.~Debbah, ``Holographic MIMO surfaces for 6G wireless networks: Opportunities, challenges, and trends," \emph{IEEE Wireless Commun.}, vol. 27, no. 5, pp. 118--125, Jul. 2020.

\bibitem{PassJamSU}
B.~Lyu, D.~T.~Hoang, S.~Gong, D.~Niyato, and D.~I.~Kim, ``IRS-based wireless jamming attacks: When jammers can attack without power," \emph{IEEE Wireless Commun. Lett.}, vol. 9, no. 10, pp. 1663--1667, Oct. 2020.

\bibitem{DaipartI}
X.~Wei, D.~Shen, and L.~Dai, ``Channel estimation for RIS assisted wireless communications: Part I-fundamentals, solutions, and future opportunities," \emph{Commun. Lett.}, vol. 25, no. 5, pp. 1398--1402, May 2021.

\bibitem{HuangCSI}
L.~Wei, C.~Huang, G.~C.~Alexandropoulos, C.~Yuen, Z.~Zhang, and M.~Debbah, ``Channel estimation for RIS-empowered multi-user MISO wireless communications," \emph{IEEE Trans. Commun.}, vol. 69, no. 6, pp. 4144--4157, Jun. 2021.

\bibitem{HuangDLRIS}
C.~Huang, R.~Mo, and C.~Yuen, ``Reconfigurable intelligent surface assisted multiuser MISO systems exploiting deep reinforcement learning," \emph{IEEE J. Sel. Areas Commun.}, vol. 38, no. 8, pp. 1839--1850, Aug. 2020.

\bibitem{AORIS}
Q.~Wu and R.~Zhang, ``Intelligent reflecting surface enhanced wireless network via joint active and passive beamforming," \emph{IEEE Trans. Wireless Commun.}, vol. 18, no. 11, pp. 5394--5409, Nov. 2019.

\bibitem{AORCG}
H.~Guo, Y.-C.~Liang, J.~Chen, and E.~G.~Larsson, ``Weighted sum-rate maximization for reconfigurable intelligent surface aided wireless networks," \emph{IEEE Trans. Wireless Commun.}, vol. 19, no. 5, pp. 3064--3076, May 2020.

\bibitem{IIRSSur}
Y.~Wang, H.~Lu, D.~Zhao, Y.~Deng, and A.~Nallanathan, ``Wireless communication in the presence of illegal reconfigurable intelligent surface: Signal leakage and interference attack," \emph{IEEE Wireless Commun.}, vol. 29, no. 3, pp. 131-138, Jun. 2022.

\bibitem{FPJDiscoIRS}
H.~Huang, Y.~Zhang, H.~ Zhang, C.~Zhang, and Z.~Han, ``Illegal intelligent reflecting surface based active channel aging: When jammer can attack without power and CSI," \emph{IEEE Trans. Veh. Technol.}, in press, Mar. 2023.

\bibitem{RISSilent}
M.~F.~Imani, D.~R.~Smith, and P.~Hougne, ``Perfect absorption in a disordered medium with programmable meta-atom inclusions," \emph{Adv. Functional Materials}, vol. 30, no. 52, 2005310, Sept. 2020.

\bibitem{PT}
X.~Zhou, B.~Maham, and A.~Hj$\rm{\phi}$rungnes, ``Pilot contamination for active eavesdropping," \emph{IEEE Trans. Wireless Commun.}, vol. 11, no. 3, pp. 903--907, Mar. 2012.

\bibitem{CSIRIS3}
H.~Guo and V.~K.~N.~Lau, ``Uplink cascaded channel estimation for intelligent reflecting surface assisted multiuser MISO systems," \emph{IEEE Trans. Signal Process.}, vol. 70, pp. 3964--3977, Jul. 2022.

\bibitem{LDetector1}
H.~Q.~Ngo, E.~G.~Larsson, and T.~L.~Marzetta, ``Energy and spectral efficiency of very large multiuser MIMO systems," \emph{IEEE Trans. Wireless Commun.} vol. 61, no. 4, pp. 1436--1449, Apr. 2013.

\bibitem{ImprefectCSI}
T.~X.~Tran and K.~C.~Teh, ``Spectral and energy efficiency analysis for SLNR precoding in massive MIMO systems with imperfect CSI," \emph{IEEE Trans. Commun.} vol. 17, no. 6, pp. 4017--4027, Jun. 2018.

\bibitem{PTDTSyn}
Q.~Yan, H.~Zeng, T.~Jiang, M.~Li, W.~Lou, and Y.~T.~Hou, ``Jamming resilient communication using MIMO interference cancellation," \emph{IEEE Trans. Inf. Forensic Secur.}, vol. 11, no. 7, pp. 1486--1499, Jul. 2016.


\bibitem{ChanAge}
K.~T.~Truong and R.~W.~Heath~Jr., ``Effects of channel aging in massive MIMO systems," \emph{J. Commun. Netw-S. Kor.}, vol. 15, no. 4, pp. 338--351, Aug. 2013.


\bibitem{Ricianfading}
J.~Zhang, L.~Dai, Z.~He, S.~Jin, and X.~Li, ``Performance analysis of mixed-ADC massive MIMO systems over Rician fading channels," \emph{IEEE J. Sel. Areas Commun.}, vol. 35, no. 6, pp. 1327--1338, Jun. 2017.



\bibitem{RCGAlg}
N.~Boumal, B.~Mishra, P.-A.~Absil, and R.~Sepulchre, ``Manopt, a MATLAB toolbox for optimization on manifolds," \emph{J. Mach. Learn. Res.}, vol. 15, no. 1, pp. 1455--1459, 2014.

\bibitem{RIS256ele}
W.~Tang, M. Z.~Chen, X.~Chen, J. Y.~Dai, Y.~Han, M. D.~Renzo, Y.~Zeng, S.~Jin, Q.~Cheng, and T. J.~Cui, ``Wireless communications with reconfigurable intelligent surface: Path loss modeling and experimental measurement," \emph{IEEE Trans. Wireless Commun.}, vol. 20, no. 1, pp. 421-439, Jan. 2021.


\bibitem{RISCB}
T.~Cui, M.~Qi, X.~Wan, J.~Zhao, and Q.~Cheng, ``Coding metamaterials, digital metamaterials and programmable metamaterials," \emph{Light-Sci. Appl.}, vol. 3, e218, Oct. 2014.

\bibitem{IRSelement}
B.~Di, H.~Zhang, L.~Song, Y.~Li, Z.~Han, and H.~V.~Poor, ``Hybrid beamforming for reconfigurable intelligent surface based multi-user communications: Achievable rates with limited discrete phase shifts," \emph{IEEE J. Sel. Areas Commun.}, vol. 38, no. 8, pp. 1809--1822, Aug. 2020.

\bibitem{IRSUPA}
B.~Di, H.~Zhang, L.~Li, L.~Song, Y.~Li, and Z.~Han, ``Practical hybrid beamforming with finite-resolution phase shifters for reconfigurable intelligent surface based multi-user communications," \emph{IEEE Trans. Veh. Technol.}, vol. 69, no. 4, pp. 4565--4570, Apr. 2020.

\bibitem{DiscreteIRS}
Q.~Wu and R.~Zhang, ``Beamforming optimization for wireless network aided by intelligent reflecting surface with discrete phase shifts," \emph{IEEE Trans. Commun.}, vol. 68, no. 3, pp. 1838--1851, Jan. 2021.

\bibitem{Frequencyhopping} D.~J.~Torrieri, ``Frequency hopping with multiple frequency-shift keying and hard decisions," \emph{IEEE Trans. Commun.}, vol. 32, no. 5, pp. 574--582, May 1984.

\bibitem{Frequencyhopping1} V.~Navda, A.~Bohra, S.~Ganguly, and D.~Rubenstein, ``Using channel hopping to increase 802.11 resilience to jamming attacks," in \emph{Proc. 26th Annu. IEEE Int. Conf. Comput. Commun.}, Anchorage, AK, USA, May 2007, pp. 2526--2530.

\bibitem{3GPP}
Further Advancements for E-UTRA Physical Layer Aspects (Release 9), document 3GPP TS 36.814, Mar. 2010.

\bibitem{Wishart}
A.~M.~Tulino, S.~Verd$\acute{\rm u}$, ``Random matrix theory and wireless communications," \emph{Foundations Trends Commun. Inf. Theory},  vol. 1, no. 1, pp. 1-182, Jun. 2004.



\end{thebibliography}
\end{document}